# Regression to the Mean can Explain Saturation of Geomagnetic Storms

**Authors:** Nithin Sivadas[1,2*], David Sibeck[2], Varsha Subramanyan[3], Maria-Theresia Walach[4], Dogacan Su Ozturk[5], Banafsheh Ferdousi[6], Bayane Michotte de Welle[2]

**Affiliations:**

[1]Department of Physics, The Catholic University of America, Washington, D.C., USA

[2] NASA Goddard Space Flight Center, Greenbelt, Maryland, USA

[3]Department of Physics, University of Illinois Urbana-Champaign, USA

[4]Department of Physics, Lancaster University, United Kingdom

[5]Department of Physics, Geophysical Institute, University of Alaska, Fairbanks, AK, USA

[6]Air Force Research Laboratory, Albuquerque, NM, USA

*Corresponding author. Email: nithinsivadas90@gmail.com

**Abstract:** Extreme space weather events on Earth occur during intervals of strong solar wind driving[1]. The solar wind drives plasma convection and currents in the near-Earth space environment[2]. For low values of the driver, the Earth's response is linear, estimated by parameters such as the polar cap index based on ground magnetometer activity[3]. Curiously, for extreme solar wind driving, the Earth's response appears not to increase beyond a saturation limit[4]. Theorists have advanced a host of explanations for this saturation effect, but there is no consensus[5]. Here, we demonstrate that this saturation is a manifestation of the *regression to the mean* effect[6] arising from random uncertainty in the timing and magnitude of solar wind measurements. Our results reveal that data analysis underpinning the saturation theories is non-linearly biased, thereby challenging the validity of the theories. Correcting for the uncertainties reveals that the Earth's response to solar wind driving is linear throughout, and that the impact of extreme geomagnetic storms can be twice as large as previously thought. We show that regression to the mean is a fundamental property of the relationship between measurement and the truth, where the truth corresponding to the measurement is closer to the mean. This effect is particularly pronounced for uncertain measurements of extreme values and is likely to manifest across various fields, from extreme climate studies to chronic medical pain.

## MAIN TEXT

A fundamental challenge in science is understanding how systems behave under rare and severe conditions. However, a critical, often-overlooked factor complicates this pursuit: uncertainty in measuring the conditions. Here, we quantitatively demonstrate that random error can systematically distort our inferences of the system's response, especially for extreme events. We resolve a long-standing puzzle in space physics—the apparent saturation of geomagnetic activity during extreme space weather[1]. We encourage readers to consult the supplementary sections in the sequence they are first referenced in the text.

**Saturation of Geomagnetic Activity.** Extreme geomagnetic storms with strong solar wind driving can increase the electric current strengths in the near-Earth space environment[7,8], causing power outages, disruptions in satellite communication, and ozone loss in the polar ionosphere[9,10]. The solar wind slows down through a bow-shock upstream of the Earth's magnetosphere and drives plasma convection in the polar cap ionosphere (See Figure 1a, and Supplementary Discussion 1.a). Early observational studies showed that the cross polar cap index (*PCI*), a measure of geomagnetic activity, increases linearly on average with the merging electric field ($E^*_m$), a measure of the solar wind driving[3] (See Extended Data Table 2 for description of symbols). The *PCI* is estimated by a polar magnetometer from the ground in each hemisphere and is proportional to the cross polar cap potential[11,12], while $E^*_m$ is measured from a spacecraft in the solar wind far upstream of the Earth. As more measurements of rare and extreme solar wind driving accumulated through the years, surprisingly, there appeared to be an upper limit to the cross-polar cap potential beyond which increasing solar wind driving led to no increase in the potential (See Figure 1b and Extended Data Fig. 6). This new inference from later measurements led to the emergence of 10 different theories attempting to explain the saturation phenomenon (See Extended Data Table 1)[4,5]. However, here we argue that there is no statistical evidence for the saturation of geomagnetic response to strong solar wind driving, and its appearance in measurements is a result of uncertainty in solar wind driver measurements.

**Uncertainty and the Problem of Definition.** Solar wind measurements are made mainly far upstream from the day-side reconnection site. They are, hence, an uncertain estimate of the actual solar wind driver close to the reconnection site in the magnetosheath, transformed during its journey, and influenced by local plasma and field conditions. Uncertainties in measurements are commonly attributed to instrumental error, but a surprising fact is that uncertainties can also result from the assumptions made while inferring data from measurements. This uncertainty stems from a "problem of definition"[13] (e.g., Supplementary Discussion 1.b).

This problem of definition arises when using solar wind measurements made at the Lagrange point L1, ~230 $R_E$ upstream of the reconnection site, as an estimate for the local magnetosheath conditions that drive the coupling of energy and plasma into the magnetosphere. The shocked solar wind driving ($E^{sh}_m$) near the reconnection site differs from what is measured at L1 [14,15] due to 1) variability in propagation times from the L1 point to the Earth, 2) spatial variability of the solar wind structure[16], and 3) evolution of solar wind along that path due to a variety of plasma processes changing the essential parameters influencing the reconnection rate, such as the local magnetic field direction.

Uncertainties can be of two types: bias or unbiased random error. A bias is a consistent deviation of the true value from the measurement itself. Physical theories that explain relations between physical parameters or their respective measurements almost always attempt to explain consistent and deterministic deviations between them, excluding the random fluctuations in their values. In contrast, the unbiased random error often consists of these random fluctuations and is the random deviation of the measurements from the true value. Physical theories in space science generally explain average changes, not random fluctuations. Hence, an unknown bias can be misinterpreted as a physical effect (i.e., an average change) when a theory is constructed to explain the bias. In short, physical theories intend to explain *physical* biases, and if we ignore bias stemming from random measurement error, we may mistake it for a physical phenomenon.

**Regression to the mean and the more-probable.** A common goal of scientific analysis is to uncover the relations between two or more physical processes. Random errors in the measurement of these processes pose a difficulty in inferring relations between them, for example, in our case, the relation between solar wind driver and the geomagnetic response. However, it is widely believed that averaging the measurements can remove the effects of random error, as the underestimates will cancel out the overestimates. Supplementary Discussion 1.c shows that this widespread belief is only partially true.

A popular technique to find the relations between two or more physical processes is regression analysis of their measurements. Here the uncertainty in the dependent variable is averaged away, but the uncertainty in the independent (or conditional) variable manifests as bias in the relation inferred between the variables, known as *regression bias*[17]. If the regression bias is non-linear, it can create an appearance of saturation of the dependent variable for increasing values of the independent variable. This statistical phenomenon is a result of a *regression to the mean*[6,18].

Supplementary Discussion 1.d explains the regression to the mean effect from five different yet consistent vantage points. For instance, in literature, regression to the mean is commonly understood as a statistical phenomenon where extreme measurements are more likely followed by measurements closer to the mean[6]. However, we argue that the phenomenon has a more fundamental origin. As discussed above it manifests as regression bias in the relation inferred between two measurements[18](See Extended Data Fig. 1b and Extended Data Fig. 5). Most importantly, it is also a probabilistic effect where the true value being measured is more likely to be closer to the mean of the stochastic process than the measurement[18] (Extended Data Fig. 1a).

The regression to the mean is not only a statistical phenomenon, though it commonly appears as one. It is not a result of a particular statistical method of inferring relations between parameters, although the particular method will be affected by it. At its core, regression to the mean is a fundamental property of the relation between the true value of a stochastic process and its measurement (See Supplementary Discussion 1.e).

The essence of regression to the mean is that the truth being measured regresses towards the mean, and this property of the relation between the truth and measurement manifests in many different forms depending on what one uses the measurements for. It is a fundamental logical consequence of the relation between the true value and its measurement, implying that for even a single measurement, the corresponding true value is likely closer to the mean of the stochastic process. For a stochastic process with a general probability distribution, the effect is more aptly named the ‘regression to the more-probable’.

**Solar wind uncertainty and error model.** Due to regression to the mean, when an extreme value of the solar wind driver is measured at the L1 Lagrange point, the true driver value near the Earth is more likely smaller and closer to the mean. Therefore, the geomagnetic response to this smaller driver will also be smaller. When we do not account for this regression to the mean, as in most previous studies, the smaller geomagnetic response will be wrongly associated with the extreme value of the solar wind driver measured at L1. The more extreme the value of the measured solar wind driver, the greater the mismatch between the true driver and the measured value. As a result, the cross-polar cap potential response to the measured uncertain solar wind driver deviates from linearity during the rare and extreme values and hence, appears to saturate.

To quantitatively test this surprising argument, we construct an error model $X^*=X(t+dt) + \epsilon$, where $X^*$ is the measured 'uncertain' solar wind driver at L1, and $X$ is the 'true' driving by the shocked solar wind close to the reconnection site. $dt$ is the uncertainty in the solar wind propagation time, and $\epsilon$ is the random magnitude changes in the driver (See details in Methods[19] and Extended Data Fig. 2). Figure 2c presents the relative uncertainty ($\epsilon(X)/X$) calculated from measurements with respect to the true shocked solar wind driver in the magnetosheath ($X$). It is at least 30%, and is heteroskedastic i.e., varying with the magnitude of the true value ($X$) [20].

When the error is heteroskedastic like some uncertainties related to the 'problem of definition', and $X$ has a log-normal distribution like many solar wind parameters[21], the likely true value for a given measurement acquires a non-linear bias mimicking a *saturation*[18] of $X$ with increasing $X^*$ (Extended Data Fig. 1c). We prove analytically that such a non-linear bias also increases with a dimensionless measure of uncertainty in time (See Supplementary Discussion 1.f, Extended Data Fig. 4c, Supplementary Methods 2.c). By incorporating the calculated uncertainties into the Monte-Carlo error model, it solves for $X^*$ and predicts an appearance of saturation of $X$ given $X^*$ (See Figure 2b). This saturated curve, and the conditional probability distribution of $X$ given $X^*$, matches surprisingly well with the apparent saturation in the data (See Figure 2a). The precondition of this non-linear saturation relation between $PCI$ and $E^*_m$ is the heteroskedastic nature of the error in $E^*_m$, the propagation time uncertainty, and the fact that the solar wind driver is a log-normal process with a particular autocorrelation time constant (See Supplementary Discussion 1.g). The error model is validated against data and robust to changes in the inputs (See Supplementary Methods 2.b and Extended Data Fig. 3).

**Saturation of the cross-polar cap potential explained.** This surprising result from the error model shows that uncertainty in the solar wind driver leads to a biased inference that the geomagnetic response saturates with solar wind driving. This apparent saturation arises when comparing the solar wind driver at L1 to the cross-polar cap index, which is proportional to the cross-polar cap potential[11]. Hence, there is currently no statistical evidence to suggest an upper limit to the energy transferred from the solar wind to the polar ionosphere. The ten prevailing theories and models of polar cap potential saturation were developed to explain this biased inference from existing erroneous measurements. As a result, these theories have not been tested or validated against corrected and unbiased data. We maintain that the original assumption —that the solar wind magnetic field lines connect directly to the polar regions, driving ionospheric convection, and therefore are on average linearly related to the magnetic fluctuations in ground-magnetometers and thus the polar cap index —holds true[22].

**Correcting the effect of uncertainties.** One way to address the effect of uncertainties is to reduce them. For instance, by relying on spacecraft measurements closer to Earth to estimate the correct driver. However, uncertainties are bound to exist in every measurement, and the problem of definition will almost always exist, so the method of reducing uncertainties is unlikely to eliminate them. Hence, alternatively, we statistically quantify the uncertainties, predict their effect, and offset them from our inference. This method is achievable, albeit challenging, through scientific investigation of the relations between the truth and measurement, and the assumptions that underlie our inferences of the measurements.

Once we calculate uncertainties and develop an error model, we can derive the non-linear bias in the regression analysis. We can subtract this bias ($b$) from the erroneous independent variable using "regression calibration"[19,20] (See Methods). Applying this bias correction to the solar wind *driver* ($E^c_m = E^*_m - b$) reveals a surprisingly linear relationship with the cross-polar cap index up to 15 mV/m – without assuming any linearity or non-linearity in the underlying relation between $E^{sh}_m$ and $PCI$, or their error model counterparts.

Figure 3a shows the saturating green curve from erroneous data transforming into the linear purple curve after regression calibration. This demonstrates that the true geomagnetic response is linear, and that the apparent saturation of the geomagnetic response results from uncertainty in the solar wind driver and the regression to the mean effect. Beyond 15 mV/m, insufficient data prevents conclusions about the shape of the relation between the driver and response. When the same calibration is applied to other measures of geomagnetic activity, such as the westward auroral electrojet (See Figure 3b and Supplementary Discussion 1.h), its relation with the driver becomes linear, providing further evidence that geomagnetic response to solar wind driving does not saturate. This implies we cannot rely on the magnetosphere to dampen the effects of extreme geomagnetic storms.

It is this linear relationship between the driver and response that a new (or old) physical theory of saturation needs to explain and validate its predictions against. The development of multiple saturation theories can be attributed to different inferences from data. Under-sampling bias, often confused with regression to the mean, leads to different saturation limits of the geomagnetic response depending on the sample size of a study (See Supplementary Discussion 1.i). Further evidence that saturation results from uncertainty in the solar wind driver is hiding in plain sight in the space science literature (See Supplementary Discussion 1.j).

**Conclusion and Generalizability.** When correctly interpreted, measurements reveal a linear rather than a saturated relation of geomagnetic response to solar wind driver on average. Thus, extreme space weather could impact Earth far more than previously thought (twice the impact at solar wind strengths extrapolated to ~25 mV/m). While saturation at unobserved solar wind strengths remain possible, there is no statistical evidence for this so far. Our result is not an alternative theory competing with the existing theories of saturation (Extended Data Table 1), but rather a demonstration that available data does not show evidence for the saturation of geomagnetic response. Hence, it challenges the foundation of existing physical theories of saturation and highlights the need to revisit and validate them with calibrated solar wind driver estimates.

In any field where uncertainty in the driver of a system is significant, and its response to extreme and rare instances of driving are of interest, the regression to the mean effect can lead to the underestimation of the system's response to extreme drivers. For example, this effect may play a role in climate models, which may underestimate extreme climate events like heatwaves[23], or the impact of strong earthquakes away from their epicenter, or severe medical symptoms and their resolution, like the effectiveness of back-pain treatments[24]. In such fields, biases and even non-linear biases may hide in plain sight and be misunderstood as a physical process.

As artificial intelligence or machine-learning models, which fall under the umbrella of non-linear, non-parametric regression analysis, become popular within physics and other fields with large quantities of measurements, the non-linear bias stemming from heteroskedastic uncertainty or temporal uncertainty is likely present. They may impact prediction and any physical inferences drawn from the model outputs, and hence we urge caution in their use and call for further study of how regression to the mean influences these models.

The insight that regression to the mean is fundamental to the relation between the truth and its measurement implies that the effect will manifest in different forms within different methods of inference. Hence, all researchers in the fields that attempt to use measurements to infer relations between the physical processes they represent need to be aware of the essence and impact of the logic of regression to the mean. Ignoring this effect can lead researchers and entire fields of inquiry down long, winding paths that deviate from the truth.

## MAIN REFERENCES

1. Baker, D. N. What is space weather? *Advances in Space Research* **22**, 7–16 (1998).

2. Dungey, J. W. Interplanetary Magnetic Field and the Auroral Zones. *Phys. Rev. Lett.* **6**, 47 (1961).

3. Troshichev, O. A. PC index as a ground-based indicator of the solar wind energy incoming into the magnetosphere: (1) relation of PC index to the solar wind electric field EKL. *Frontiers in Astronomy and Space Sciences* vol. 9 Preprint at https://doi.org/10.3389/fspas.2022.1069470 (2022).

4. Shepherd, S. G. Polar cap potential saturation: Observations, theory, and modeling. *J. Atmos. Sol. Terr. Phys.* **69**, 234–248 (2007).

5. Borovsky, J. E., Lavraud, B. & Kuznetsova, M. M. Polar cap potential saturation, dayside reconnection, and changes to the magnetosphere. *J. Geophys. Res. Space Phys.* **114**, doi:10.1029/2009JA014058 (2009).

6. Barnett, A. G., van der Pols, J. C. & Dobson, A. J. Regression to the mean: what it is and how to deal with it. *Int. J. Epidemiol.* **34**, 215–220 (2005).

7. Gonzalez, W. D. *et al.* What is a geomagnetic storm? *J. Geophys. Res. Space Phys.* **99**, 5771–5792 (1994).

8. Coxon, J. C. *et al.* Extreme Birkeland Currents Are More Likely During Geomagnetic Storms on the Dayside of the Earth. *J. Geophys. Res. Space Phys.* **128**, e2023JA031946 (2023).

9. Eastwood, J. P. *et al.* The Economic Impact of Space Weather: Where Do We Stand? *Risk Analysis* **37**, 206–218 (2017).

10. National Research Council. *Severe Space Weather Events: Understanding Societal and Economic Impacts: A Workshop Report. Severe Space Weather Events--Understanding Societal and Economic Impacts* (National Academies Press, Washington, DC, 2008). doi:10.17226/12507.

11. Troshichev, O., Hayakawa, H., Matsuoka, A., Mukai, T. & Tsuruda, K. Cross polar cap diameter and voltage as a function of PC index and interplanetary quantities . *J. Geophys. Res. Space Phys.* **101**, (1996).

12. Lockwood, M. & McWilliams, K. A. A Survey of 25 Years' Transpolar Voltage Data From the SuperDARN Radar Network and the Expanding-Contracting Polar Cap Model. *J. Geophys. Res. Space Phys.* **126**, e2021JA029554 (2021).

13. Taylor, J. R. (John R. *An Introduction to Error Analysis : The Study of Uncertainties in Physical Measurements*. (University Science Books, 1982).

14. Walsh, B. M., Bhakyapaibul, T. & Zou, Y. Quantifying the Uncertainty of Using Solar Wind Measurements for Geospace Inputs. *J. Geophys. Res. Space Phys.* **124**, 3291–3302 (2019).

15. Morley, S. K., Welling, D. T. & Woodroffe, J. R. Perturbed Input Ensemble Modeling With the Space Weather Modeling Framework. *Space Weather* **16**, 1330–1347 (2018).

16. Milan, S. E., Carter, J. A., Bower, G. E., Fleetham, A. L. & Anderson, B. J. Influence of Off-Sun-Earth Line Distance on the Accuracy of L1 Solar Wind Monitoring. *J. Geophys. Res. Space Phys.* **127**, e2021JA030212 (2022).

17. Borovsky, J. E. Noise, Regression Dilution Bias, and Solar-Wind/Magnetosphere Coupling Studies. *Frontiers in Astronomy and Space Sciences* **0**, 45 (2022).

18. Sivadas, N. & Sibeck, D. G. Regression Bias in Using Solar Wind Measurements. *Frontiers in Astronomy and Space Sciences* **0**, 159 (2022).

19. Methods are available at the end of this Main Text.

20. Carroll, R. J., Ruppert, D., Stefanski, L. A. & Crainiceanu, C. M. Measurement error in nonlinear models: A modern perspective, second edition. *Measurement Error in Nonlinear Models: A Modern Perspective, Second Edition* 1–455 (2006).

20. Veselovsky, I. S., Dmitriev, A. V. & Suvorova, A. V. Lognormal, Normal and Other Distributions Produced by Algebraic Operations in the Solar Wind. *AIP Conf. Proc.* **1216**, 152 (2010).

22. Vasyliunas, V. M. Concepts of Magnetospheric Convection. in *The Magnetospheres of the Earth and Jupiter* (ed. Formisano, V.) 179–188 (Springer Netherlands, Dordrecht, 1975).

23. Kornhuber, K., Bartusek, S., Seager, R., Schellnhuber, H. J. & Ting, M. Global emergence of regional heatwave hotspots outpaces climate model simulations. *Proceedings of the National Academy of Sciences* **121**, e2411258121 (2024).

24. Whitney, C. W. & Von Korff, M. Regression to the mean in treated versus untreated chronic pain. *Pain* **50**, 281–285 (1992).

# FIGURE LEGENDS

**Figure 1| Saturation of geomagnetic activity with solar wind driving. a,** Solar wind merging electric field ($E^*_m$) measured upstream of the bow-shock, is transformed to the shocked solar wind driver in the magnetosheath ($E^{sh}_m$) downstream of the bowshock, which corresponds to the dawn-dusk electric field that maps to the polar ionosphere ($E_{PC}$); **b,** Observations from 1995-2019, green curve, shows that on average the cross polar cap index ($\propto$ the dawn-dusk electric field $E_{PC}$) increases linearly at low values of solar wind driving, but deviates from linearity and saturates at large values of driving. The purple curve shows that the error model developed in this work predicts the same saturation effect due to uncertainty in the solar wind driver instead of any physical mechanism. The transparent shaded area for both curves shows a 95% confidence interval.

**Figure 2| Regression bias predicted by the error model matches saturation effect from data. a,** Conditional probability density function of 25 years of measurements $PCI|E^*_m \sim E_{PC}|E^*_m$; **b,** Conditional probability density function reproduced by the error model pdf($X|X^*$) which is very similar to the measurements presented in panel a. The transparent shaded area for both regression curves shows a 95% confidence interval; **c,** Quantified relative uncertainty $\epsilon(X)/X$ due to random variations in solar wind driver measured at L1 ($E^*_m$, $X^*$) for a measured value of the shocked solar wind driver ($E^{sh}_m$, $X$).

**Figure 3| Correcting the effect of random errors in solar wind driver values reveals a linear geomagnetic response. a,** Polar cap index varies linearly with solar wind driving on average (purple curve), after correcting the regression bias in the erroneous solar wind driver values (green curve). **b,** The westward auroral electrojet strength also varies linearly with solar wind driving
288 ($E^c_m$) on average (purple curve) after the same regression bias correction applied to the erroneous solar wind driver values ($E^*_m$) (green curve). The transparent shaded area for all the curves shows a 95% confidence interval.

**Extended Data Figure 1| Explaining regression to the mean effect using probability theory and regression bias. a,** For a physical parameter $X$ which has an underlying probability distribution, some values are more likely than others, and hence when the measurement is $X^*=3$ the corresponding true value is more likely to be $X=2$ than $X=3$. **b,** The expected true value of a Gaussian process $X$ has a linear bias when measurement is $X^* = X + \epsilon$ with a constant random error ($\epsilon$). **c,** The expected true value of a log-normal process $X$ has a non-linear bias when the measurement $X^*$ has random error $\epsilon(X)$ that varies with the strength of $X$. This figure is adapted from Sivadas & Sibeck (2022)[18], used under CC BY 4.0. The transparent shaded area for regression curves shows a 95% confidence interval.

**Extended Data Figure 2| Stochastic properties of inputs to the error model. a,** The black line shows the pdf of the polar cap index $E_{PC}$ from 1995 to 2019. And the purple line shows the lognormal pdf of the model random variable $X$ that best fits the pdf of $E_{PC}$. **b,** The black dashed line shows the autocorrelation function of the polar cap index $E_{PC}$ averaged over data from 1995 to 2019. And the purple line shows the autocorrelation function of the model random variable $X$ that closely follows the ACF of $E_{PC}$. **c**, The black line shows the uncertainty in the propagation delay from L1 to the bow-shock nose ($dt_1$). The blue line shows the uncertainty in the propagation delay from the bow-shock nose to the polar cap ionosphere ($dt_2$). The purple line shows the total uncertainty in the propagation delay from L1 to the polar cap ($dt_1 + dt_2$).

**Extended Data Figure 3| Statistics of the output of the error model. a,** The model predicts with good correspondence the pdf of the estimate of the merging electric field propagated to the polar cap, with a probability density difference of less than ~0.1 between data and model for most values. $E^*_m$ is data, and $X^*$ is model. **b,** The model predicts the nature of the random error between the estimate and the true value of the merging electric field projected on the ionosphere. The black line is the normalized uncertainty associated with $E^*_m$ (data), and the purple line is the normalized uncertainty associated with $X^*$ (model). **c,** There is a non-linear conditional bias in $E^*_m$:$\langle E^*_m - E_{PC}|E_{PC}\rangle$ that varies with $E_{PC}$, which is reproduced very well by the model $\langle X^* - X|X\rangle$. **d,** Conditional normalized error distribution from data, **e,** Conditional normalized error distribution from the model. Both data and model show similar trends in how the error varies with increasing strength of the true variable –a steep increase in the spread of the error from 0 to ~12 mV/m in the vertical axis, and then a decrease (or loss of power) beyond that point.

**Extended Data Figure 4| Sensitivity analysis of magnitude uncertainty. a,** Non-linear regression bias caused by heteroskedastic magnitude uncertainty increases with a measure of the uncertainty represented by the minimum relative magnitude uncertainty. **b,** The heteroskedastic variation of the magnitude uncertainty with the true value $X$ is shown, for different values of the minimum relative magnitude uncertainty corresponding to the different regression bias curves in panel a. **c,** Non-linear bias caused by uncertainty in time increases with a measure of the uncertainty represented by the dimensionless ratio of the time uncertainty ($\delta$) and the autocorrelation constant ($k$).

**Extended Data Figure 5| Another graphical explanation of the regression to the mean effect.** (Left) Probability distribution function of the true variable $X$, with each bin colored uniquely. (Right) The probability distribution function of the inaccurate estimate of $X$, i.e., $X^*$ with the color representing the magnitude of $X$. The amount of a particular color per bin represents the proportion of $X$ values misidentified as $X^*$.

**Extended Data Figure 6| Increasing trend of the saturation limits observed by previous observational studies. a,** Cross polar cap potential response to solar wind driving is linear up to a saturation limit, which varies depending on the observational study and the methodology they use. All except Boyle et al., 1997 [25] infer a saturation beyond the limit in $E^*_m$. See the studies here: 1 [26], 2 [27], 3 [28], 4 [25], 5 [29], 6 [30], 7 [31], 8 [32], 9 [33]. **b,** The saturation limit of different studies presented here follows an increasing trend with the year of study (Pearson correlation coefficient ~0.61), with the latest studies including higher number of data samples.

**Extended Data Table 1| Theories of polar cap potential saturation.** Ten models explaining cross polar cap potential saturation. See the following works for more details about each of the models: 1 [34,35], 2 [36,37] , 3 [38,39], 4 [40], 5 [39,41], 6 [42], 7 [43], 8 [44], 9 [45],10 [46,47] and an assessment of most of them[5]. Adapted with permission from [Joseph E. Borovsky, Benoit Lavraud, Maria M. Kuznetsova, Table A1, Polar cap potential saturation, dayside reconnection, and changes to the magnetosphere., Journal of Geophysical Research: Space Science, Publisher: John Wiley and Sons] Copyright 2009 by American Geophysical Union.

**Extended Data Table 2| Common symbols used in this work.** Variables representing measurements and their corresponding counter parts in the error model used commonly through the paper are summarized here.

**Fig. 1| Saturation of geomagnetic activity with solar wind driving.**

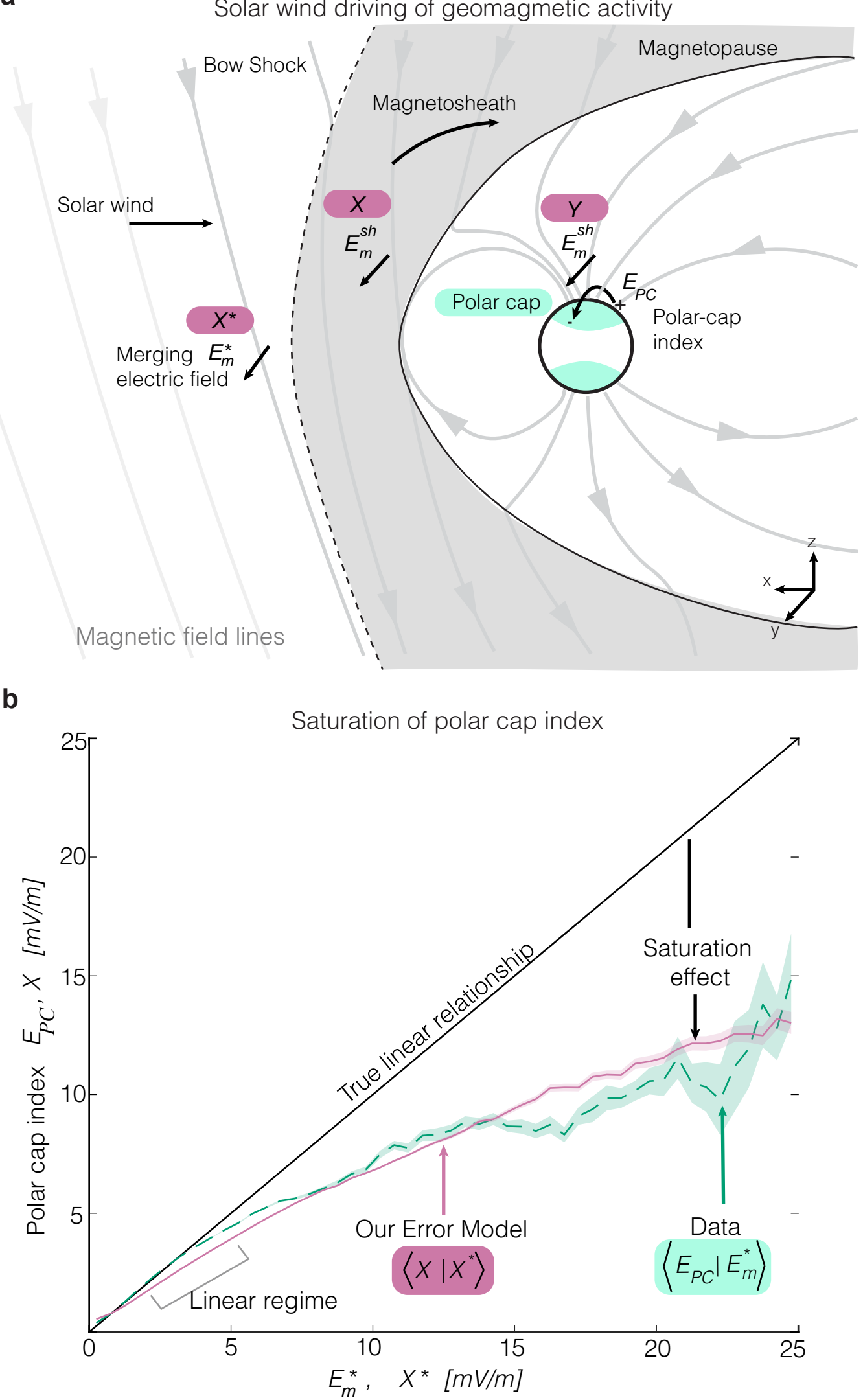

**Fig. 2| Regression bias predicted by the error model matches saturation effect from data..**

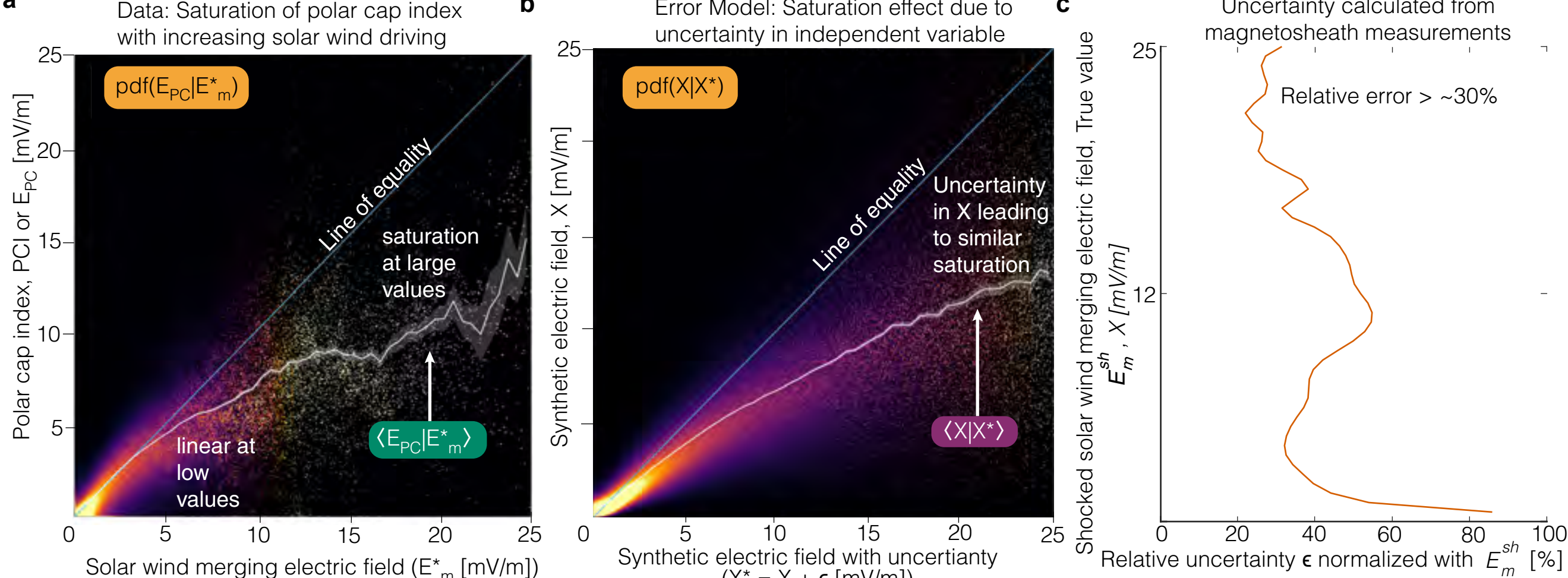

**Fig. 3| Correcting the effect of random errors in solar wind driver values reveals a linear geomagnetic response.**

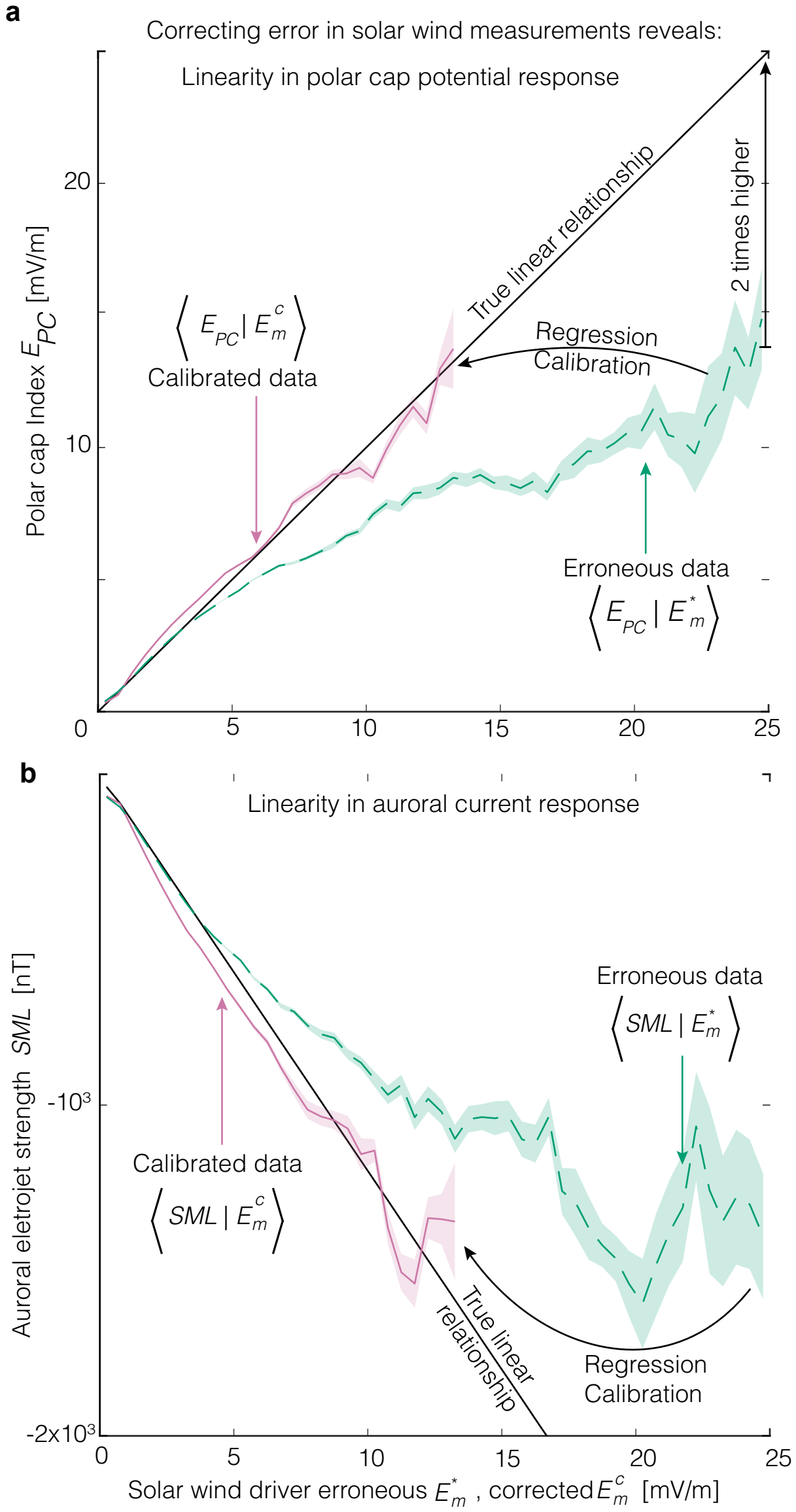

## Extended Data Fig.1| Explaining regression to the mean effect using probability theory and regression bias.

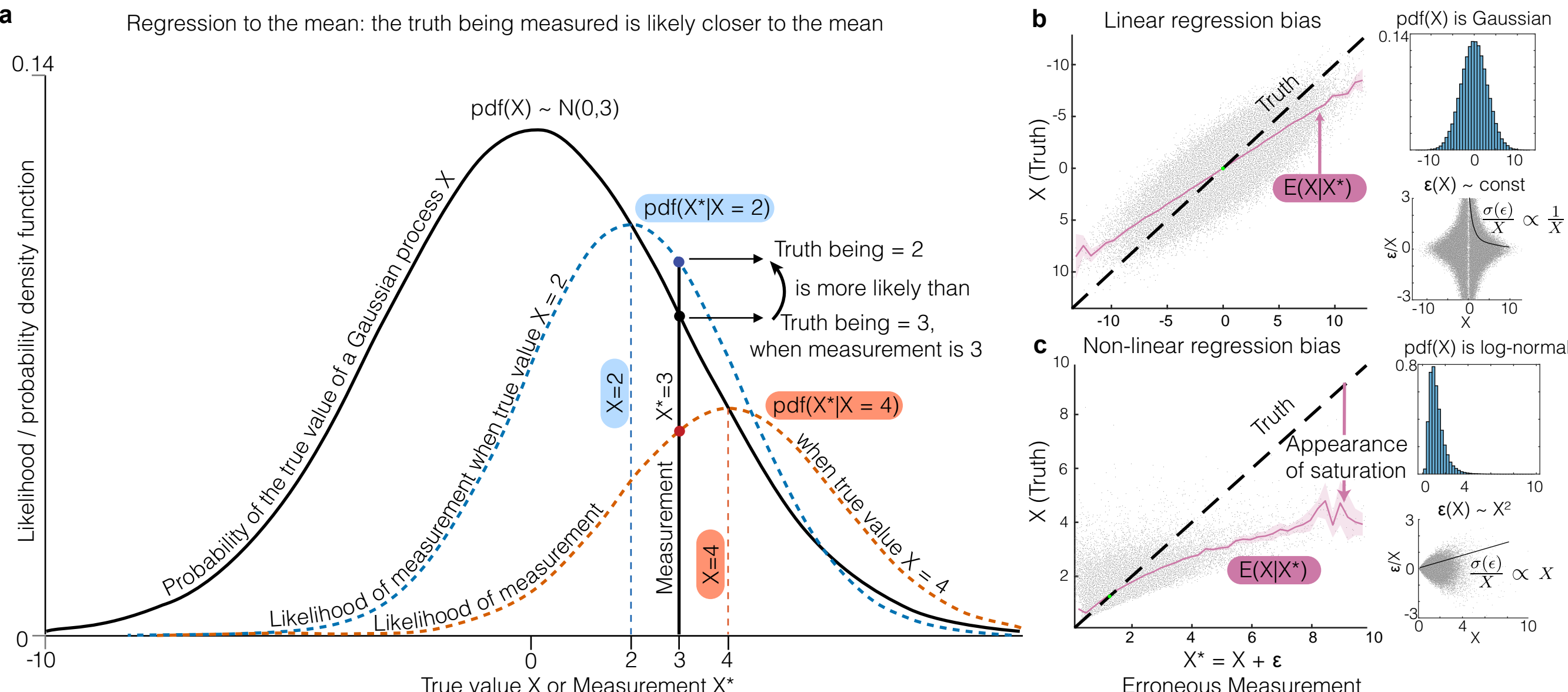

# Extended Data Fig.2| Stochastic properties of inputs to the error model.

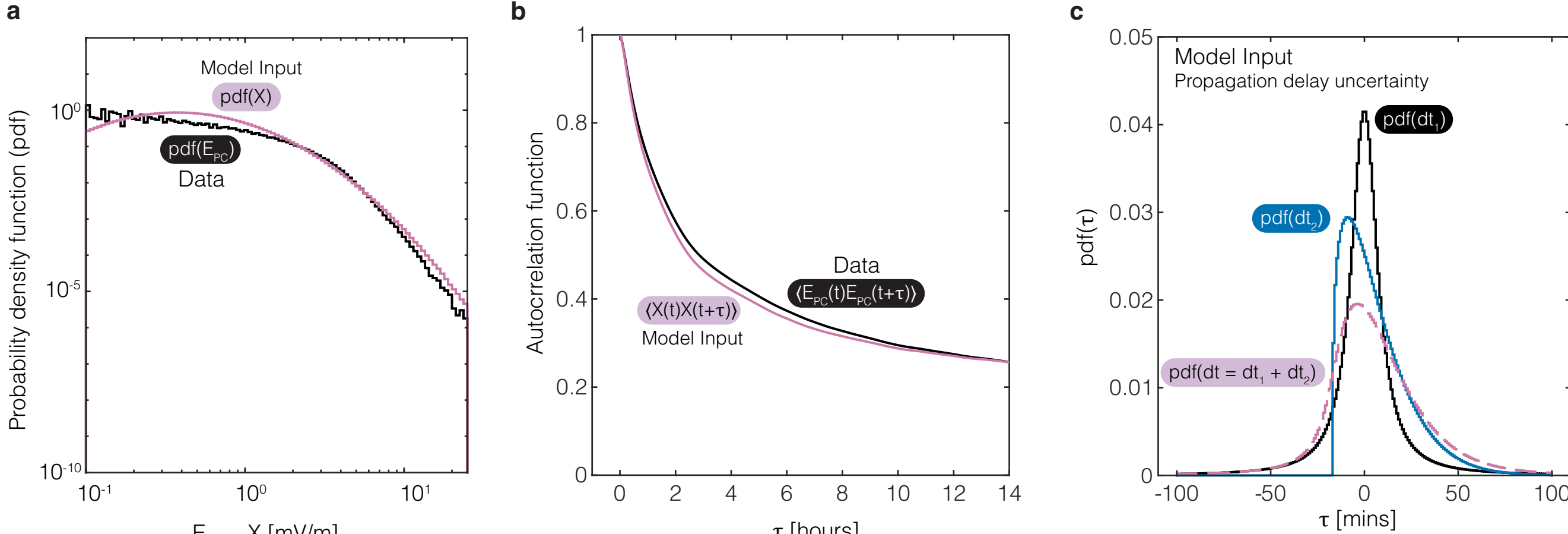

# Extended Data Fig.3| Statistics of the output of the error model.

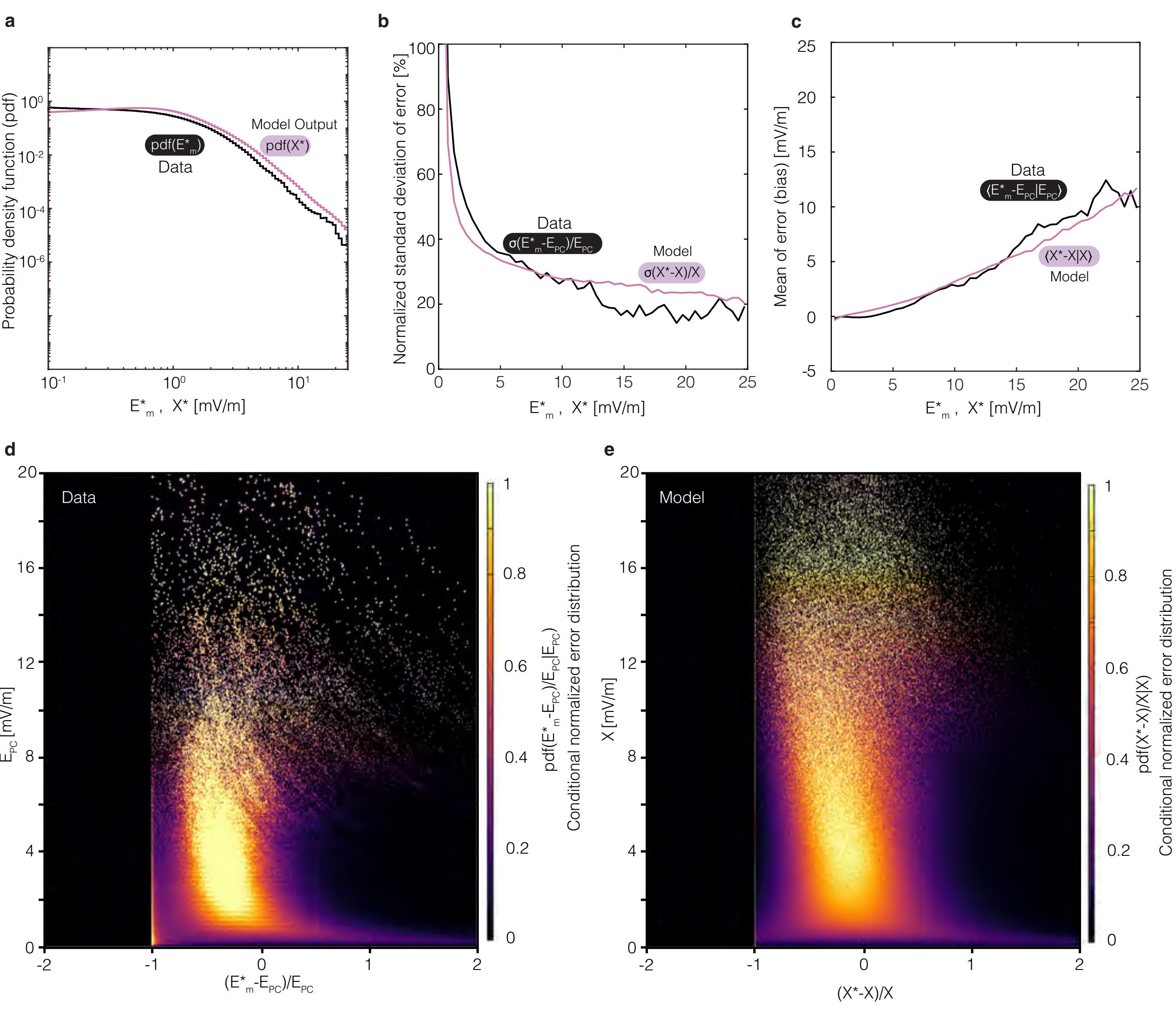

## Extended Data Fig.4| Sensitivity analysis of magnitude uncertainty

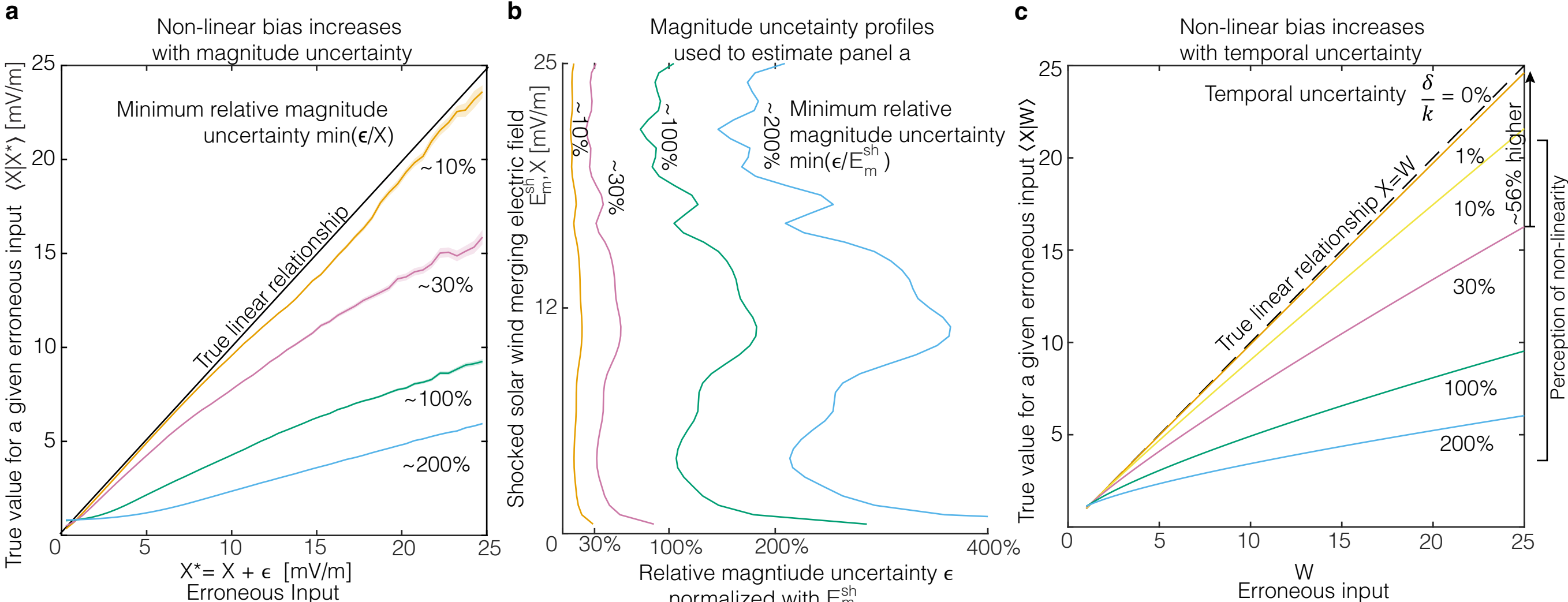

# Extended Data Fig.5| Another graphical explanation of the regression to the mean effect

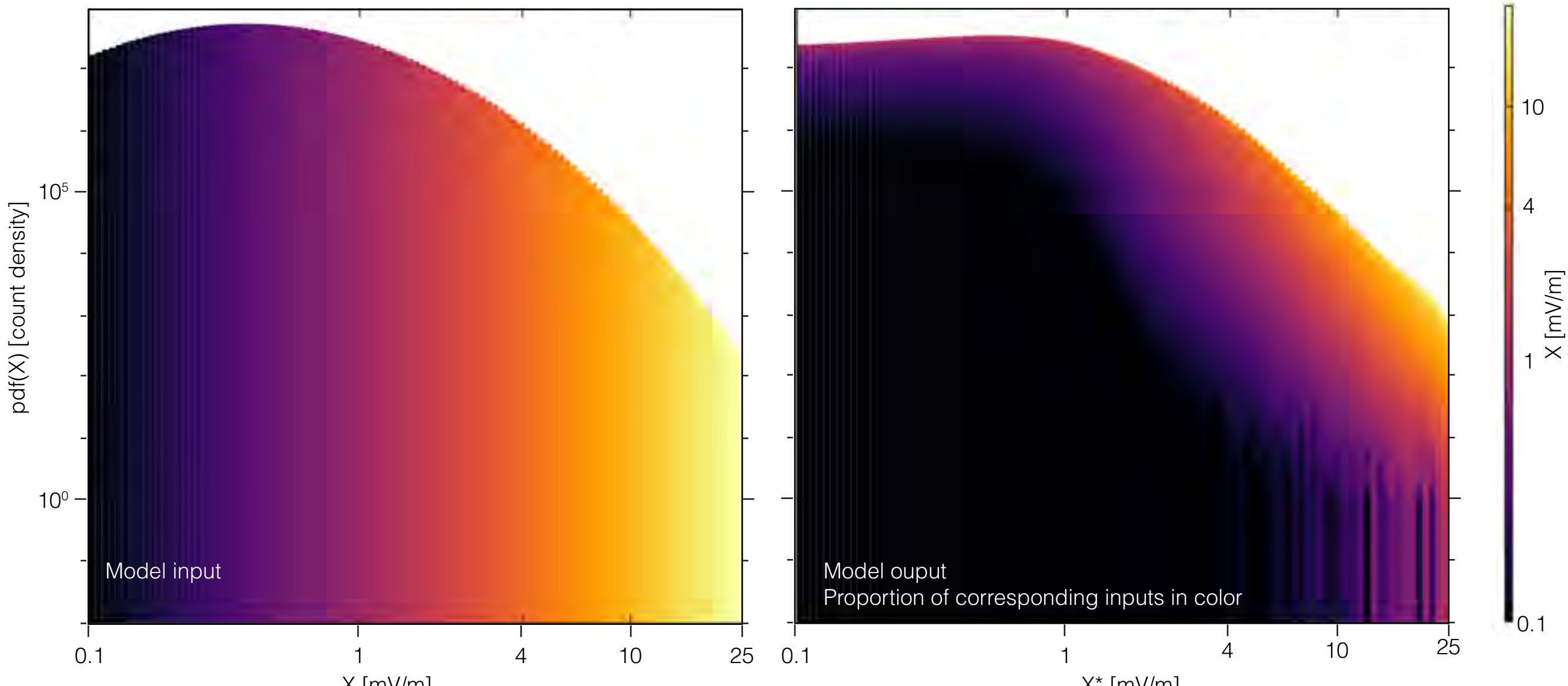

**Extended Data Fig.6| Increasing trend of the saturation limits observed by previous observational studies**

**a**

| Study | Saturation limit in $E_m^*$ [mV/m] | Instruments |
|---|---|---|
| 1) Reiff et al., 1981 | ~ 3 | LEO Satellites: AE-C, AE-D, S3-3 |
| 2) Wygant et al., 1983 | ~ 0.5 | LEO Satellites: S3-3 |
| 3) Weimer et al., 1990 | ~ 4 | Magnetometers, AE Index |
| 4) Boyle et al., 1997 | ~ 8 (found no saturation) | LEO Satellites: DMSP |
| 5) Russell et al., 2001 | ~ 3 | AMIE |
| 6) Liemohn & Ridley, 2002 | ~ 10 | AMIE |
| 7) Shepherd et al., 2002 | ~ 3 | SuperDARN radars |
| 8) Nagatsuma, 2002 | ~ 5 | Magnetometer PC Index |
| 9) Hairston et al., 2005 | ~ 10 | LEO Satellites: DMSP |

**b**

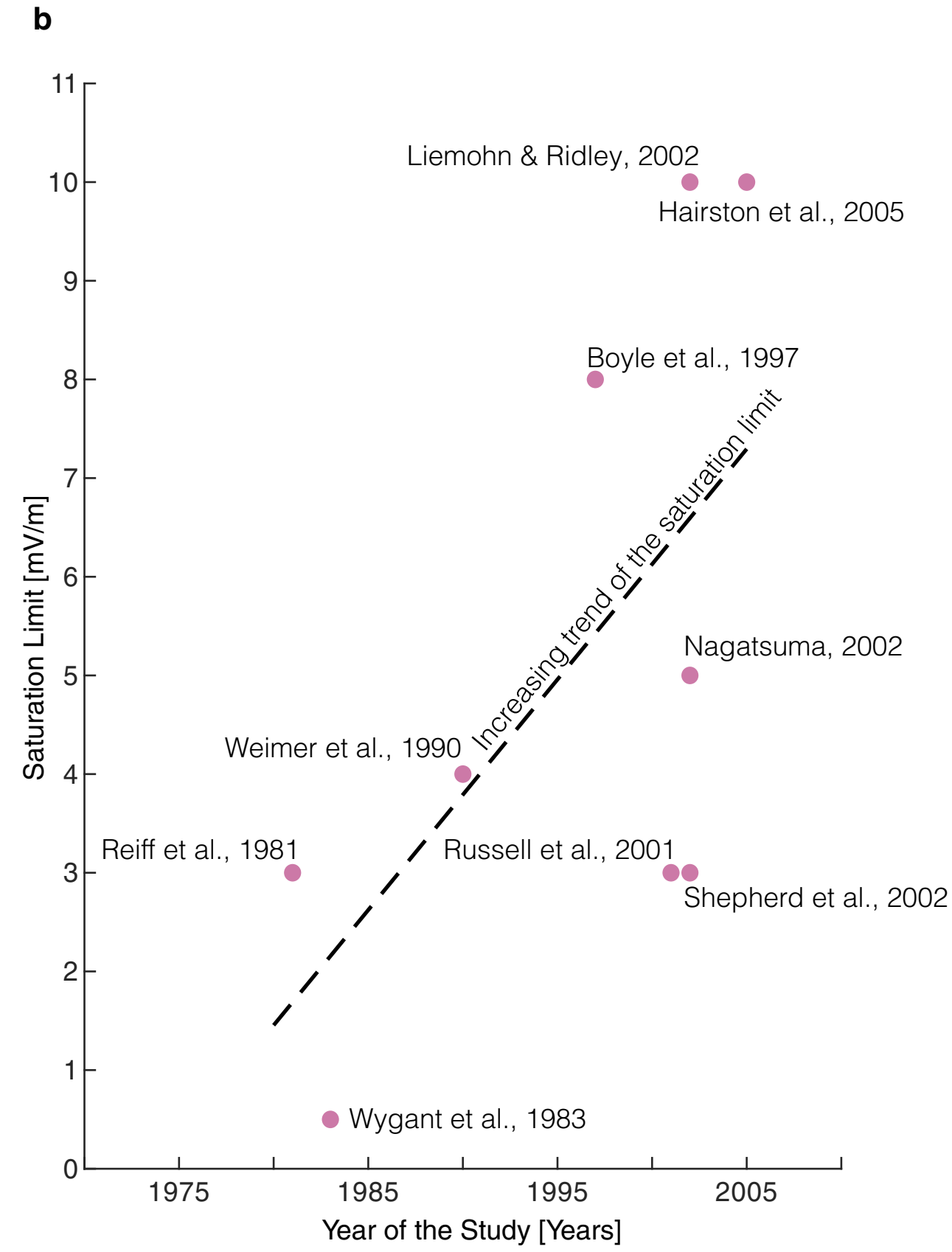

## Extended Data Table 1| Theories of polar cap potential saturation

| Theory/Model | Description |
|---|---|
| 1) Hill model | Reduced dayside magnetosheath field lowers the reconnection rate |
| 2) Flaring model | Flaring lowers the reconnection rate |
| 3) X-line length model | Magnetosphere size changes decreases total reconnection |
| 4) Force-balance model | Magnetosheath flow divergence reduces geoeffective length |
| 5) Magnetosheath flow model | Solar wind conditions alter dayside reconnection rate |
| 6) Lobe-bulge model | Bulging lobes impedes dayside reconnection |
| 7) Bow shock model | Polar cap currents alter bow shock and dayside reconnection rate |
| 8) Ionospheric outflow model | Ionospheric outflow mass-loads magnetospheric convection |
| 9) Ram pressure-saturation model | Dynamic pressure constraints currents |
| 10) MHD generator model | Alfven conductance of the shocked solar wind plasma limits currents |

## Extended Data Table 2| Common symbols used in this work

| Symbols | Description | Symbols | Description |
|---|---|---|---|
| $E_m^*$ | Solar wind driver, Kan-Lee electric field, merging electric field, measured at L1 Lagrange point and time-shifted to bow-shock | $X^*$ | Measurement, Model counter part of solar wind driver measured at L1 Lagrange point and time-shifted to bow-shock |
| $E_m^{sh}$ | Shocked solar wind driver measured in the magnetosheath close to the subsolar point near the magnetopause | $X$ | True value, Model counter part of shocked solar wind driver close to the subsolar point in the magnetosheath |
| $E_{sw}$ | Solar wind convection electric field | $\epsilon$ | Zero-mean random error, magnitude uncertainty |
| $E_{PC}$ | Dawn-dusk convection electric field across the polar cap | $Y_{PC}$ | Model counterpart of dawn-dusk convection electric field across the polar cap, also counter part of $PCI$ |
| $E_m^c$ | Calibrated solar wind driver | $X^c$ | Calibrated solar wind driver |
| $E_R$ | Reconnection electric field | $\boldsymbol{b}$ | Regression bias (due to regression to the mean effect) |
| $PCI$ | Polar cap index, also a proxy for $E_{PC}$ | $Y$ | True value, model counter part of geomagnetic response |
| $PCN$ | Polar cap index derived from a magnetometer in northern polar cap | $Y^*$ | Erroneous measurement of $Y$ |
| $PCS$ | Polar cap index derived from a magnetometer in the southern polar cap | $dt$ | Uncertainty in the time of measurement |
| $SML$ | Westward auroral electrojet index derived from SuperMAG network of ground magnetometers | $dt_1$ | Uncertainty in solar wind propagation delay from L1 to bow shock nose |
| $B_{T,sw}$ | Transverse magnitude of solar wind magnetic field | $dt_2$ | Uncertainty in propagation of the effect of solar wind from bow shock to the polar cap ionosphere |
| $B_z$ | Magnetic field along the Z-axis of the GSM coordinates | $\sigma_\delta$ | Standard deviation of random uncertainty in time of measurement |
| $B_{sh}$ | Magnetic field value in the magnetosheath | $k$ | Autocorrelation time constant |
| $V_{sh}$ | Plasma velocity measured in the magnetosheath | $V_{sw}$ | Solar wind plasma velocity |
| $\theta'$ | Shear angle between the magnetic field vector in the magnetosheath and the magnetosphere at the magnetopause | $\theta_{sw}$ | Solar wind magnetic field clock angle |

# METHODS

The most common symbols used are explained in the Extended Data Table 2.

## Solar wind measurement uncertainty

There are several solar wind coupling functions that attempt to quantify the driving, one important one is the solar wind merging electric field or merging geoeffective field $E^*_m$[4,48].

$$E^*_m = V_{sw} B_{T,sw} \sin^2(\theta_{sw}/2) \qquad [1.1]$$

This field is also known as the Kan-Lee electric field and is calculated from solar wind parameters at Lagrange point L1 alone[48]. In equation 1.1, $V_{sw}$ is the solar wind speed in km/s, $B_{T,sw} = (B^2_y + B^2_z)^{1/2}$ is the transverse magnitude of the interplanetary magnetic field (IMF) in nT, and $\theta_{sw} = \tan^{-1}(B_y/B_z)$ is the transverse IMF clock angle in radians. $E_m$ is a positive-valued quantity with the units of the electric field [mV/m], reflecting the assumption that energy flows only from the solar wind to the magnetosphere. The above quantities are measured in Geocentric Solar Magnetospheric (GSM) coordinates using spacecrafts orbiting the L1 point ~230 earth Radii ($R_E$) upstream from Earth. The GSM coordinates are convenient for studying the effects of the IMF components on magnetospheric and ionospheric phenomena. Its X-axis points towards the Sun, and the Z-axis is the projection of the Earth's magnetic dipole axis onto the plane perpendicular to the X-axis. For purely southward IMF, the merging electric field $E^*_m$ is equal to the solar wind electric field $E_{sw} = V_{x,sw}B_z$, where $V_{x,sw}$ is a component of the solar wind velocity along the x-direction.

The dawn-dusk portion of the shocked solar wind convection electric field maps along the equipotential magnetic field lines and drives the plasma convection down in the polar cap ionosphere. The convection corresponds to an electric field across the polar cap in the rest-frame of the Earth ($E_{PC}$). The polar cap ($PC$) index is a measure of this field and part of the energy input into the Earth's magnetosphere. The index has gone through many iterations over the past 50 years, but it is essentially the maximum amplitude of variations observed from magnetometers near the south and north pole[3,49,50]. The current version of the index is scaled on a statistical basis of magnetic variations to the merging electric field, such that the index is highly correlated to the merging electric field. This makes the $PC$ index independent of daily and seasonal variations and the local ionospheric properties. $PCN$ is the index derived from the magnetometer in the northern polar cap, while $PCS$ is derived from the southern polar cap. We use the $PCI$ index as a proxy for $E_{PC}$ (i.e. $PCI \sim E_{PC}$), as it combines both the indices to provide a positive-valued index, which is more accurate than the individual indices[50]. $PCI$=0.5($PCS$+$PCN$), with a condition that negative values of either $PCS$ or $PCN$ are set to 0. $PCI$ or $E_{PC}$ also has the units of the electric field, and several statistical analyses in literature reveal that it is linearly proportional on average to the electric potential across the polar cap, called the cross-polar cap potential[11,51]. Fluctuations in $PCI$ can occur due to night-side magnetosphere processes[52], but the above relation is still maintained on average.

The Kan-Lee electric field $E^*_m$ is only an approximation of the true driver of the magnetosphere i.e., the shocked solar wind plasma $E^{sh}_m = V_{sh}B_{sh}\sin^2(\theta'/2)$, whose value is not easily available to us (See Supplementary Methods 2.a). This leads to the uncertainty and non-linear regression bias discussed in this work. Instead, we only have an erroneous estimate of the true coupling function approximately propagated to the polar cap, $E^*_m$. Hence, we need to reinterpret the literature as saying that low values of the erroneous estimate of the solar wind strengths $E^*_m$ (not $E^{sh}_m$) correlate linearly with the polar cap potential. And at high values of $E^*_m$, the polar cap potential saturates.

We calculate $E^*_m$ by using WIND satellite measurements published in the OMNIWeb database[53]. The database provides the values corrected for the propagation delay of the solar wind from L1 to the bow-shock nose [54–56]. Then as commonly done, we apply a further correction of a constant delay of ~17 minutes to account for the propagation delay from the nose to the polar cap ionosphere[16,57]. Previous literature that discusses the polar cap potential saturation problem estimates $E^*_m$ similarly. We use the WIND data from 1995 to 2019 and the polar cap indices from the same time range. Both data are 1-minute averages, larger time averages will lead to additional uncertainties[58]. We only use data samples when both WIND measurements and polar cap indices are available.

$E^*_m$ can differ from the shocked solar wind driver in the magnetosheath $E^{sh}_m$ in two ways. One is through a consistent deterministic bias, that is determined by the bow-shock that slows down the plasma. This deterministic bias is nearly zero as the tangential (and the largest) component of the electric field across the bow-shock remains unchanged. We do not concern ourselves with this deterministic bias, as it will manifest in the data as a physical effect anyway. However, $E^*_m$ can also differ randomly from the true solar wind driver $E^{sh}_m$, due to fluctuations in the plasma properties caused by physical processes between L1 and the reconnection site, acting in random directions. These random fluctuations contribute to the uncertainty in $E^*_m$ when we use it as a proxy for the true driver $E^{sh}_m$. This random error is concerning, as it does not average away, and manifests as a regression bias in data analysis and is easily confused as a physical effect, hence we calculate and correct for it.

We categorize the random uncertainty in $E^*_m$ relative to the true value $E^{sh}_m$ into three primary sources:

1. Uncertainty in propagation delay of the solar wind from L1 to the bow-shock nose ($dt_1$)
2. Uncertainty in the propagation delay of the effect of solar wind forcing from bow-shock nose to the polar cap ionosphere ($dt_2$)
3. Random variability in the shocked solar wind due to spatial variation in the incoming solar wind and transformations of its plasma parameters during propagation through the bow-shock and the magnetosheath ($\epsilon$)

Knowing the statistical distribution of the above uncertainty will allow us to construct a stochastic model of the estimate $E^*_m$ as a function of the true driver $E^{sh}_m$ mapped to the polar cap ionosphere. In the following subsection, we develop a statistical error model of the estimate of the true shocked solar wind that drives the cross-polar cap convection from an estimate of its stochastic properties and uncertainty distributions. For the calculated uncertainties (Figure 2c), the model predicts the polar cap potential saturation, which is strikingly similar to data from 25 years of observations (see Figure 1b and Figure 2a-b).

**Statistical error model**

To distinguish between data and model, we will replace $E^{sh}_m$ with $X$ when referring to the random variable corresponding to the shocked solar wind driver in the statistical model. We also replace the polar cap index $PCI$, which is proportional to the convection electric field $E_{PC}$ in the ionosphere with a counterpart in the model $Y_{PC}$. Hence, in the model, $X$ is the shocked solar wind driver accurately time-shifted to the polar cap, and $X^*$ is its erroneous estimate (i.e. solar wind driver measured at L1 or $E^*_m$). We hypothesize the following statistical error model:

$$X^*(t) = X(t+dt_1+dt_2) + \epsilon(t) \qquad [1.2]$$

We assume that $X^*$, $X$, $dt_1$, $dt_2$, and $\epsilon$ are stochastic processes. In other words, they are each a collection of random variables in time ($t \in T$) with an associated probability distribution that determines the random value it might take at a given time $t$. Below we present our estimates of the probability distributions and autocorrelation functions of these stochastic processes. Using these estimates of $X$, $dt_1$, $dt_2$, and $\epsilon$ we calculate $X^*$. After which, we validate the model results by comparing them with the data $E^*_m$.

**Input *X***: If reconnected open-field lines in the polar cap region do not have large parallel resistances, accurate mapping of the shocked solar wind electric field onto the polar cap should result in $X$, such that $Y_{PC} \propto X$. In agreement with this, in data, we see that, the probability density function (pdf) of the *PCI* index or $E_{PC}$ (and hence $Y_{PC}$) and solar wind driver $E^*_m$ are very similar, implying the distribution of $Y_{PC}$ is also similar to $X$. As a result, we assume that the pdf of $X$ is similar to the pdf of $E_{PC}$ (and $E^*_m$) that we can estimate from data. Note that this is not surprising since $E_{PC}$ is constructed to be highly correlated with $E^*_m$, and from Kan and Lee's arguments $E^*_m$ is also correlated with $E^{sh}_m$, and hence all these parameters have similar probability distribution functions. (Though we have measurements of $E^{sh}_m$ from near-Earth orbiting spacecraft, they are discontinuous and hence do not directly provide an unbiased pdf of $X$.) Like many solar wind parameters, $E_{PC}$ and $E^*_m$ can be approximated to be a lognormal distribution[21]. Therefore, we assume the pdf of model input $X$ to be a lognormal distribution that closely fits the pdf of $E_{PC}$ from data (See Extended Data Fig. 2a). We also assume the adjacent values of $X$ in time are correlated similarly to adjacent values of $E_{PC}$ in time (See Extended Data Fig. 2b), as fluctuations in time of shocked solar wind electric field should correlate with that of the cross-polar cap electric field in the ionosphere. In other words, we assume that in our model, $X$ shares the stochastic properties of the polar cap index or electric field $E_{PC}$. If the assumption fails, the results of the model, particularly the second order statistics, will be inconsistent with what is observed in the data. Finally, we assume $X$ is a stationary process, i.e., its probability distribution and autocorrelation function does not change with time.

**Uncertainty in propagation delay from L1 to Nose $dt_1$**: Solar wind parameters measured upstream are time-shifted to account for the delay in propagation of the wind to the bow-shock nose. Case and Wild estimate the uncertainty to be on the order of minutes [59]. Based on their results and consistent with other studies[60], we assume $dt_1$ to be an independent random process with a pdf of a Student's t-distribution with shape factor 1.3, mean 0, and scale parameter ~8 minutes (See Extended Data Fig. 2c and Supplementary Methods 2.b). The distribution is zero-mean and has a longer tail than the normal distribution.

**Uncertainty in propagation delay from nose to polar cap $dt_2$**: Changes in the day-side shocked solar wind electric field propagate along equipotential open magnetic field lines to the polar caps. The delay in this propagation is on average ~17 minutes[16] but can vary from -5 to 50 minutes[51]. Historically, researchers have used a constant propagation delay ($t_2$) of about ~10-30 minutes[28,33] for this stage of propagation. However, the uncertainty of propagation time here is significant. We model the pdf of $t_2$ as a Weibull distribution with mean ~ 17 minutes[16], standard deviation ~ 25 minutes, and shape factor 1.3 (See Extended Data Fig. 1c and Supplementary Methods 2.b). The Weibull distribution keeps the total delay $t_2+dt_2$ positive and captures the broad spread in the propagation delay[16,50,51].

**Magnitude uncertainty $\epsilon$**: There are several other reasons for the magnitude of the shocked solar wind driver to be randomly different from its proxy measured at the L1. The IMF clock angle could change substantially after crossing the bow shock, spatial variations in the solar wind can lead to a different part of the wind interacting with the Earth's magnetosphere, and changes in the magnetospheric state or its history can lead to changes in the local plasma and field conditions in magnetosheath. As the solar wind strength increases, the random variation in the field can increase due to the increased spatial structuring of the solar wind, and any clock angle variation during increased field strength can lead to larger variations in geoeffectiveness.

To estimate these random variations, we use direct evidence from a database of simultaneous measurements of plasma and field strengths in the magnetosheath. Using a gradient boost classification algorithm, measurements from near-Earth satellites such as THEMIS, MMS, DoubleStar, and Cluster, are classified into solar wind, magnetosheath, and magnetosphere regions[61]. Our interest is in the measurements made within the magnetosheath, close to the magnetopause in the subsolar region ($|Y|<5$ $R_E$, and $|Z| < 5$ $R_E$). The distance from the magnetopause boundary is determined using a machine-learning based empirical model of the boundary, which performs better than other magnetopause models available in literature[62]. For specific values of the magnetosheath measurements ($E^{sh}_m$), we calculate the variance in the measurements made at L1 and time-shifted to the bow-shock ($E^*_m$). This variance is an estimate of the uncertainty in the solar wind driver $E^*_m$ for a given value of the shocked solar wind driver $E^{sh}_m$. The "magnitude uncertainty" $\epsilon$ is set to the same variance, and we model it as a zero-mean Gaussian with a standard deviation that varies with the magnitude of $X$ according to data. The magnitude uncertainty is not constant with the strength of the shocked solar wind driver, instead it increases until $E^{sh}_m$ ~12 mV/m, after which the statistics become poor and an accurate estimate of the variance becomes challenging (See Supplementary Methods 2.b). In this regime, we estimate the uncertainty in the solar wind driver relative to measurements of near-Earth satellites just upstream of the bow-shock (~$10<X<50$ $R_E$). This uncertainty will be the minimum uncertainty for values >12 mV/m (i.e., a conservative estimate of the uncertainty), and this remains roughly the same with increasing shocked solar wind driver value. Figure 2c shows the relative error, i.e. $\sigma(\epsilon)/X$, and how that varies with $X$. The 'heteroskedastic' behavior of this uncertainty is consistent with the observed statistical variations in the difference between solar wind driver and the polar cap index shown in Extended Data Fig. 3d-e. The variation in their difference increases up to a polar cap index of ~12 mV/m and then remains constant. We also assume that $\epsilon$ has an autocorrelation function similar to that of the difference between the observed solar wind driver $E^*_m$ and the polar cap index $PCI$~$E_{PC}$.

Using the above estimates of uncertainties, we generate an ensemble of time series of $X$, $dt_1$, $dt_2$, and $\epsilon$, and calculate the corresponding time series of the erroneous estimate $X^*$ using equation [1.2]. The statistical properties of the model output agree remarkably with that of the data.

## Validation of Error Model

We validate this non-linear error model by comparing the second order statistics of the error model parameters with that of their counterparts in data. The error model predicts the pdf of the solar wind driver $E^*_m$, the standard deviation of the normalized error, the conditional normalized error distribution, and finally the regression bias stemming from the regression to the mean effect. This validation is the rationale for using the error model and is presented in the Supplementary Methods 2.b. Here we also discuss a sensitivity analysis of the non-linear regression bias, which increases with increasing uncertainty in the time of propagation, and with increasing uncertainty in the magnitude uncertainty $\epsilon$ (See Extended Data Fig. 4). The analysis shows that output of the error model is robust to the input uncertainties and variance of input $X$. Supplementary Methods 2.d summarizes the full computer code used.

## Correcting the error

The regression bias $b = X^* - \langle X|X^*\rangle$, which is the consistent deviation between the measurement and the likely true value given the measurement. The erroneous measurement $X^*$ can now be corrected to $X^c$ by subtracting the bias from it $X^c = X^* - b$, which simply reduces to $X^c = \langle X|X^*\rangle$. Crucially, from the model output, we can calculate the conditional expectation of the true value given the erroneous estimate: $f_r(x) = \langle X|X^*= x\rangle$, i.e., we have a statistical, functional relationship between the true value and the erroneous measurement. For the particular assumptions of uncertainties in measurement, $f_r(x)$ is the best estimate of the true value ($X^c$ or $E^c_m$) given the erroneous value ($X^*$ or $E^*_m$). Therefore, $f_r(x)$ can be used to statistically correct the erroneous estimate $E^*_m$ to a most likely estimate of the true value $E^c_m$. $E^c_m$ is useful for revealing the true statistical correlation with other variables, like the polar cap index $E_{PC}$ or auroral current strengths $SML$. The process of statistically correcting for the uncertainty using the conditional expectation of the true value given the erroneous estimate: $E^c_m= f_r(E^*_m)$ for an unbiased regression analysis is called regression calibration[20]. Figure 3 shows the results of this calibration. $SML$ is the westward auroral electrojet index from the SuperMAG database, which is a measure of the westward currents flowing in the auroral regions[63,64]. It is calculated using ground magnetometers and procedures completely independent of the calculation of the polar cap index. Since regression calibration of $E^*_m$ to $E^c_m$, results in a linear relationship between both $E^c_m$ and $E_{PC}$ (Figure 3a) and $E^c_m$ and $SML$ (Figure 3b), it suggests our findings are independent of the construction of the ionospheric response variables. The confidence intervals shown in Figures 1-3 only provide an estimate of the reliability of the regression procedure, i.e., it means 95% probability that repeating the procedure will contain the regression curve and does not imply that the true value or the true regression curve will lie within the confidence interval with a 95% probability.

## Temporal uncertainty can lead to a non-linear regression function

In Supplementary Methods 2.c we analytically derive the following general result that temporal uncertainty in measurement can lead to a perception of non-linearity in a linear system's correlation response. For a linear system with response $Y$, input $X$, and the input with measurement error $W=X(t + \Delta)$, the biased system response $\langle Y|W\rangle$ is a function $f_r^*(w, \sigma_\delta/k)$. Here $\sigma_\delta$ is a measure of the random uncertainty $\Delta$ in the time of measurement. And $k$ is the autocorrelation time constant. The system response is a function of the ratio $\sigma_\delta/k$. This ratio is a measure of uncertainty in the time of measurement. We numerically integrate the function, which is fully described in equation [1.12] in the Supplementary Methods 2.c, within the ranges of $\Delta$ for specific values of $\sigma_\delta/k$ and generate Extended Data Fig. 4c. The figure shows that $f_r^*(w, \sigma_\delta/k)$ varies non-linearly with $w$ when there is temporal uncertainty.

## ADDITIONAL REFERENCES

25. Boyle, C. B., Reiff, P. H. & Hairston, M. R. Empirical polar cap potentials. *J. Geophys. Res. Space Phys.* **102**, 111–125 (1997).

26. Reiff, P. H., Spiro, R. W. & Hill, T. W. Dependence of polarcap potential drop on interplanetary parameters. *J. Geophys. Res.* **86**, 7639–7648 (1981).

27. Wygant, J. R., Torbert, R. B. & Mozer, F. S. Comparison of S3-3 polar cap potential drops with the interplanetary magnetic field and models of magnetopause reconnection. *J. Geophys. Res. Space Phys.* **88**, 5727–5735 (1983).

28. Weimer, D. R., Reinleitner, L. A., Kan, J. R., Zhu, L. & Akasofu, S.-I. Saturation of the auroral electrojet current and the polar cap potential. *J. Geophys. Res. Space Phys.* **95**, 18981–18987 (1990).

29. Russell, C. T., Luhmann, J. G. & Lu, G. Nonlinear response of the polar ionosphere to large values of the interplanetary electric field. *J. Geophys. Res. Space Phys.* **106**, 18495–18504 (2001).

30. Liemohn, M. W. & Ridley, A. J. Comment on 'Nonlinear response of the polar ionosphere to large values of the interplanetary electric field' by C. T. Russell et al. *Journal of Geophysical Research: Space Physics* vol. 107 SIA 13-1 Preprint at https://doi.org/10.1029/2002JA009440 (2002).

31. Shepherd, S. G., Greenwald, R. A. & Ruohoniemi, J. M. Cross polar cap potentials measured with Super Dual Auroral Radar Network during quasi-steady solar wind and interplanetary magnetic field conditions. *J. Geophys. Res. Space Phys.* **107**, SMP 5-1 (2002).

32. Nagatsuma, T. Saturation of polar cap potential by intense solar wind electric fields. *Geophys. Res. Lett.* **29**, 62–1 (2002).

33. Hairston, M. R., Drake, K. A. & Skoug, R. Saturation of the ionospheric polar cap potential during the October–November 2003 superstorms. *J. Geophys. Res. Space Phys.* **110**, 9–26 (2005).

34. Hill, T. W., Dessler, A. J. & Wolf, R. A. Mercury and Mars: The role of ionospheric conductivity in the acceleration of magnetospheric particles. *Geophys. Res. Lett.* **3**, 429–432 (1976).

35. Siscoe, G. L. *et al.* Hill model of transpolar potential saturation: Comparisons with MHD simulations. *J. Geophys. Res. Space Phys.* **107**, (2002).

36. Merkin, V. G. *et al.* Global MHD simulations of the strongly driven magnetosphere: Modeling of the transpolar potential saturation. *J. Geophys. Res. Space Phys.* **110**, (2005).

37. Merkin, V. G. *et al.* Relationship between the ionospheric conductance, field aligned current, and magnetopause geometry: Global MHD simulations. *Planet. Space Sci.* **53**, 873–879 (2005).

38. Raeder, J. & Lu, G. Polar cap potential saturation during large geomagnetic storms. *Advances in Space Research* **36**, 1804–1808 (2005).

39. Ridley, A. J. A new formulation for the ionospheric cross polar cap potential including saturation effects. *Ann. Geophys.* **23**, 3533–3547 (2005).

40. Lopez, R. E. *et al.* Role of magnetosheath force balance in regulating the dayside reconnection potential. *J. Geophys. Res. Space Phys.* **115**, 12216 (2010).

41. Pudovkin, M. I., Zaitseva, S. A., Bazhenova, T. A. & Andrezen, V. G. Electric fields and currents in the earth's polar caps. *Planet. Space Sci.* **33**, 407–414 (1985).

42. Raeder, J., Wang, Y. L., Fuller-Rowell, T. J. & Singer, H. J. Global Simulation of Magnetospheric Space Weather Effects of the Bastille Day Storm. *Solar Physics 2001 204:1* **204**, 323–337 (2001).

43. Hernandez, S., Lopez, R. E., Wiltberger, M. & Lyon, J. G. MHD Simulations of Solar Wind-Geospace Coupling. in *AGU Spring Meeting Abstracts* SM51C-03 (SAO/NASA Astrophysics Data System, 2007).

44. Winglee, R. M., Chua, D., Brittnacher, M., Parks, G. K. & Lu, G. Global impact of ionospheric outflows on the dynamics of the magnetosphere and cross-polar cap potential. *J. Geophys. Res.* **107**, (2002).

45. Siscoe, G. L., Crooker, N. U. & Siebert, K. D. Transpolar potential saturation: Roles of region 1 current system and solar wind ram pressure. *J. Geophys. Res. Space Phys.* **107**, (2002).

46. Ridley, A. J. Alfvén wings at Earth's magnetosphere under strong interplanetary magnetic fields. *Ann. Geophys.* **25**, 533–542 (2007).

47. Kivelson, M. G. & Ridley, A. J. Saturation of the polar cap potential: Inference from Alfvén wing arguments. *J. Geophys. Res. Space Phys.* **113**, 5214 (2008).

48. Kan, J. R. & Lee, L. C. Energy coupling function and solar wind-magnetosphere dynamo. *Geophys. Res. Lett.* **6**, 577–580 (1979).

49. Troshichev, O. & Sormakov, D. PC index as a proxy of the solar wind energy that entered into the magnetosphere: 2. Relation to the interplanetary electric field e KL before substorm onset. *Earth, Planets and Space* **67**, 1–11 (2015).

50. Stauning, P. The polar cap (PC) index combination, PCC: Relations to solar wind properties and global magnetic disturbances. *Journal of Space Weather and Space Climate* **11**, 19 (2021).

51. Stauning, P. The Polar Cap PC Indices: Relations to Solar Wind and Global Disturbances. in *Exploring the Solar Wind* (InTech, 2012). doi:10.5772/37359.

52. Lockwood, M., Cowley, S. W. H. & Freeman, M. P. The excitation of plasma convection in the high-latitude ionosphere. *J. Geophys. Res. Space Phys.* **95**, 7961–7972 (1990).

53. King, J. H. & Papitashvili, N. E. Solar wind spatial scales in and comparisons of hourly Wind and ACE plasma and magnetic field data. *J. Geophys. Res. Space Phys.* **110**, 2104 (2005).

54. Weimer, D. R. *et al.* Predicting interplanetary magnetic field (IMF) propagation delay times using the minimum variance technique. *J. Geophys. Res. Space Phys.* **108**, 1026 (2003).

55. Haaland, S., Paschmann, G. & Sonnerup, B. U. Ö. Comment on "A new interpretation of Weimer et al.'s solar wind propagation delay technique" by Bargatze et al. *J. Geophys. Res. Space Phys.* **111**, 6102 (2006).

56. Knetter, T., Neubauer, F. M., Horbury, T. & Balogh, A. Four-point discontinuity observations using Cluster magnetic field data: A statistical survey. *J. Geophys. Res. Space Phys.* **109**, 6102 (2004).

57. Troshichev, O. A., Janzhura, A. S. & Stauning, P. Magnetic activity in the polar caps: Relation to sudden changes in the solar wind dynamic pressure. *J. Geophys. Res. Space Phys.* **112**, 11202 (2007).

58. Laundal, K. M. *et al.* Time-scale dependence of solar wind-based regression models of ionospheric electrodynamics. *Scientific Reports 2020 10:1* **10**, 1–9 (2020).

59. Case, N. A. & Wild, J. A. A statistical comparison of solar wind propagation delays derived from multispacecraft techniques. *J. Geophys. Res. Space Phys.* **117**, 2101 (2012).

60. Tasnim, S. *et al.* Estimation and assessment of solar wind propagation time from the Lagrange point L1 to Earth's bow shock. *Frontiers in Astronomy and Space Sciences* **12**, 1675769 (2025).

61. Nguyen, G. *et al.* Massive Multi-Mission Statistical Study and Analytical Modeling of the Earth's Magnetopause: 1. A Gradient Boosting Based Automatic Detection of Near-Earth Regions. *J. Geophys. Res. Space Phys.* **127**, (2022).

62. Michotte de Welle, B. *et al.* Global Environmental Constraints on Magnetic Reconnection at the Magnetopause From In Situ Measurements. *J. Geophys. Res. Space Phys.* **129**, e2023JA032098 (2024).

63. Gjerloev, J. W. The SuperMAG data processing technique. *J. Geophys. Res. Space Phys.* **117**, 9213 (2012).

64. Newell, P. T. & Gjerloev, J. W. Evaluation of SuperMAG auroral electrojet indices as indicators of substorms and auroral power. *J. Geophys. Res. Space Phys.* **116**, (2011).

65. Lepping, R. P. *et al.* The WIND magnetic field investigation. *Space Science Reviews 1995 71:1* **71**, 207–229 (1995).

66. Sivadas, N. *et al.* Simultaneous measurements of solar wind from L1, shocked solar wind from Magnetosheath, and measures of geomagnetic response from ground-based indices . Preprint at https://doi.org/10.5281/zenodo.17546718 (2026).

67. Sivadas, N. Code for Regression to the Mean of Extreme Geomagnetic Storms . Preprint at https://doi.org/10.5281/zenodo. 17559999 (2026).

68. Cowley, S. ~W. ~H. & Lockwood, M. Excitation and decay of solar wind-driven flows in the magnetosphere-ionosphere system. *Ann. Geophys.* **10**, 103–115 (1992).

69. Lockwood, M. Solar Wind—Magnetosphere Coupling Functions: Pitfalls, Limitations, and Applications. *Space Weather* **20**, e2021SW002989 (2022).

70. Troshichev, O. A. & Andrezen, V. G. The relationship between interplanetary quantities and magnetic activity in the southern polar cap. *Planet. Space Sci.* **33**, (1985).

71. Dunham, D. ~W. ISEE-3, the first libration-point satellite. in *Bulletin of the American Astronomical Society* vol. 11 805–806 (1979).

72. Bonilla, M. G., Mark, R. K. & Lienkaemper, J. J. Statistical relations among earthquake magnitude, surface rupture length, and surface fault displacement. *Bulletin of the Seismological Society of America* **74**, 2379–2411 (1984).

73. Borovsky, J. E. On the Saturation (or Not) of Geomagnetic Indices. *Frontiers in Astronomy and Space Sciences* **8**, 175 (2021).

74. Pulkkinen, T. I. *et al.* Magnetosheath control of solar wind-magnetosphere coupling efficiency. *J. Geophys. Res. Space Phys.* **121**, 8728–8739 (2016).

75. Doyle, M. A. & Burke, W. J. S3-2 measurements of the polar cap potential. *J. Geophys. Res. Space Phys.* **88**, 9125–9133 (1983).

76. Barnett, A. G., van der Pols, J. C. & Dobson, A. J. Regression to the mean: what it is and how to deal with it. *Int. J. Epidemiol.* **34**, 215–220 (2005).

77. Carroll, R. J., Ruppert, D., Stefanski, L. A. & Crainiceanu, C. M. Measurement error in nonlinear models: A modern perspective, second edition. *Measurement Error in Nonlinear Models: A Modern Perspective, Second Edition* 1–455 (2006).

78. Vörös, Z. *et al.* Probability density functions for the variable solar wind near the solar cycle minimum. *J. Geophys. Res. Space Phys.* **120**, 6152–6166 (2015).

79. Borovsky, J. E. Noise and Solar-Wind/Magnetosphere Coupling Studies: Data. *Frontiers in Astronomy and Space Sciences* **0**, (2022).

80. Michotte de Welle, B., aunai, . nicolas ., Nguyen, G., Ghisalberti, A., Connor, H., Génot, V., Jeandet, A., & Lavraud, B. (2026). MANGO Dataset : Magnetosphere Atlas from Normalized Geospace Observations [Data set]. *Zenodo*. https://doi.org/10.5281/zenodo.19671706

## Data Availability

All data we have used is publicly available. We thank GSFC/SPDF OMNIWeb service for the WIND spacecraft measurements[65]. The WIND spacecraft measurements of the solar wind propagated to the bow-shock can be accessed from https://spdf.gsfc.nasa.gov/pub/data/omni/high_res_omni/sc_specific/. The MMS, Cluster, DoubleStar, and THEMIS data can be accessed from GSFC/SPDF web service https://spdf.gsfc.nasa.gov/. The polar cap index values PCN and PCS were downloaded from https://pcindex.org/archive. We thank Dr. Oleg Troshichev of the Arctic and Antarctic Research Institute for this data (https://pcindex.org/contacts). The auroral electrojet indices were downloaded from the SuperMAG database https://supermag.jhuapl.edu/indices/. For SuperMAG indices we gratefully acknowledge the SuperMAG collaborators (https://supermag.jhuapl.edu/info/?page=acknowledgement). This Zenodo repository archives the data used in the manuscript[66] (https://doi.org/10.5281/zenodo.17546718). The broader dataset of magnetosheath measurements is published here (https://doi.org/10.5281/zenodo.19671706)[80].

## Code Availability

All the relevant data and the MATLAB and Python codes to process and visualize them have been curated and uploaded in the associated compute capsule and archived here[67] (https://doi.org/10.5281/zenodo.17559999).

## Acknowledgements

We thank the International Space Science Institute (ISSI) for supporting meetings in Bern, through the International Team project #25-654: “Why does the Geomagnetic Response to the Solar Wind Saturate?”.

## Funding Statement

N.S. discloses support for the research of this work from NASA Heliophysics Guest Investigator Program Grant # 80NSSC23K0446 and NSF AGS GEM Award #2453576. D.S. discloses support of the research of this work from NASA Heliophysics Participating Investigator Program Grant #WBS516741.01.24.01.03. V.S. declares no relevant research funding. M.T.W. discloses support for the research of this work from UKRI STFC through the Ernest Rutherford Fellowship (ST/X003663/1). D.S.O discloses support for the research of this work from NSF AGS GEM

Award #2453576. B.F. declares no relevant research funding. B.M.W discloses support for this work from NASA Postdoctoral Program at NASA Goddard Space Flight Center, administered by Oak Ridge Associated Universities under contract with NASA.

## Author contributions

N.S. constructed the error model, carried out the associated data analysis, and implemented the regression bias correction method. V.S., N.S., derived analytically the effect of time uncertainty on the regression function. N.S., M.T.W., B.M.W, developed and reviewed the computer code and the compute capsule. N.S., M.T.W., validated the error model using second-order statistics. D.S.O, N.S., carried out the literature survey of polar cap saturation studies. N.S., D.S.O, B.F., D.S., led the funding acquisition. N.S. led the conceptualization with input from D.S., V.S., M.T.W. N.S., D.S., supervised and administered the research. N.S., V.S., D.S., wrote the original draft, which was revised and edited by N.S. with feedback from all authors.

## Competing interests

Authors declare that they have no competing interests.

## Additional Information

Supplementary Information is available for this paper.

Correspondence and requests for materials should be addressed to Nithin Sivadas, nithinsivadas90@gmail.com.

Reprints and permissions information is available at www.nature.com/reprints.

The views expressed are those of the author and do not necessarily reflect the official policy or position of the Department of the Air Force, the Department of Defense, or the U.S. government. Approved for public release #AFRL20262399.

# Supplementary Information Guide for

# Regression to the Mean can Explain Saturation of Geomagnetic Storms

**Authors:** Nithin Sivadas[1,2*], David Sibeck[2], Varsha Subramanyan[3], Maria-Theresia Walach[4], Dogacan Su Ozturk[5], Banafsheh Ferdousi[6], Bayane Michotte de Welle[2]

**Affiliations:**

[1]Department of Physics, The Catholic University of America, Washington, D.C., USA

[2] NASA Goddard Space Flight Center, Greenbelt, Maryland, USA

[3]Department of Physics, University of Illinois Urbana-Champaign, USA

[4]Department of Physics, Lancaster University, United Kingdom

[5]Department of Physics, Geophysical Institute, University of Alaska, Fairbanks, AK, USA

[6]Air Force Research Laboratory, Albuquerque, NM, USA

*Corresponding author. Email: nithinsivadas90@gmail.com

**Supplementary Information** contains **Supplementary Discussion 1a-1j** and **Supplementary Methods 2a-2d**. The Supplementary Discussion explains the space physics background, regression to the mean as a fundamental property, the interpretation of the error model, and the linearity of the geomagnetic response after correcting for regression bias. Supplementary Methods describe the utility of the Kan-Lee electric field as a driver function, validation of the error model, the analytical derivation of how time uncertainty causes regression bias, and a summary of the data analysis procedure used in the computer code.

# 1 Supplementary Discussion

### a. Physics of solar-wind magnetosphere coupling

The Earth's magnetic field carves out a cavity in the oncoming solar wind, called the magnetosphere. It mostly keeps out the plasma and magnetic field associated with the solar wind except under specific local magnetic field conditions that lead to the reconnection of solar wind and magnetospheric magnetic field lines. Upstream from the reconnection-site, the supersonic and superalfvénic solar wind slows down through a bowshock, forming the magnetosheath just upstream from the magnetosphere's outer boundary—the magnetopause. When the shocked solar wind magnetic fields lie anti-parallel to those of the Earth, the field can undergo a topological change that opens up the Earth's magnetic field lines and reconnects them to the shocked solar wind magnetic field lines[2]. Through reconnection, the solar wind plasma and electric field can enter the Earth's magnetosphere and drive plasma convection in the magnetosphere. Conveyed to the polar ionosphere closer to the Earth, this convection electric field applies an electric potential across the polar cap on average[69] (See Figure 1a). The projection of the solar wind convection electric field along the dawn-dusk direction in the polar cap is the Kan-Lee or merging electric field ($E_m^*$)[48], which we refer to as solar wind driving[70].

In the 1970s, estimates of the cross-polar cap potential (and its corresponding dawn-dusk electric field) were made from single ground magnetometers situated in the northern and southern polar regions. The argument was that the perturbations of the magnetic field measured by the polar magnetometers give an excellent continuous estimate of the solar wind driving, as the polar regions magnetically connect to the solar wind[71]. Continuous estimates of solar wind driving from satellite missions were not yet available at the time[72]. As the magnetized plasma in the solar wind and magnetosphere is collisionless, the resistance along magnetic field lines is minuscule. Hence, the magnetohydrodynamic theory is valid, ensuring that plasma along an entire field-line moves as one[2]. Therefore, the solar wind electric field, resulting from the motion of the wind, at one end of the magnetic field line, would map along the field line to the polar ionosphere. This electric field then drives currents in the ionosphere that cause magnetic perturbations measured from ground magnetometers. Supporting this theory, the cross polar cap index ($PCI$), a measure of geomagnetic activity, increases linearly on average with the merging electric field ($E_m^*$), a measure of the solar wind driving[3]. The $PCI$ is estimated by a polar magnetometer from the ground in each hemisphere and is proportional to the cross polar cap potential[11,12], while $E_m^*$ is measured from a spacecraft in the solar wind far upstream of the Earth. Through the years, as more measurements of rare and extreme solar wind driving accumulated, the correlation between the geomagnetic response and the solar wind driver began to deviate from the linear relation. There appeared to be an upper limit to the cross-polar cap potential with increasing driving (See Figure 1b green curve).

The new inference that the cross-polar cap potential saturates with increasing solar wind driving led to the emergence of 10 different but sometimes overlapping theories and models attempting to explain the phenomenon (See Extended Data Table 1)[4,5]. The theories fall into two categories, one arguing that the conditions at the solar wind/magnetosphere interaction region (i.e., the dayside magnetosheath) during extreme solar wind driving lead to a consistent diminishing of the rate of reconnection or energy coupling from the solar wind to the magnetosphere. The other set of theories argues that processes within the magnetosphere slow down ionospheric plasma convection during extreme solar wind driving. All theories suggest that the energy transferred into the polar ionosphere during extreme space weather has, in essence, an upper limit. Hence, the theories imply that the Earth's magnetosphere shields us from extreme geomagnetic storms by inhibiting energy transfer from the strong solar wind to the ionosphere, one way or the other.

In this paper, we present a radically different proposition. We argue that there is no statistical evidence for the saturation, and its appearance in measurements is a result of uncertainty in the driver. We correct for the effect of this uncertainty and demonstrate that the geomagnetic response varies linearly with solar wind driving, implying that the energy transfer from extreme solar wind conditions to the polar ionosphere can be far greater than currently believed. The importance of this work goes beyond space science to any correlation study between a measure of a system's input and its response, revealing a larger response for extreme input values.

### b. An example of the problem of definition

A carpenter might measure the height of a door frame using a measuring tape. However, the door she creates using this height will likely not fit the frame. Even though the least count of the tape is small and does not contribute much to the uncertainty, there is uncertainty due to an implicit assumption that the height of the door is a one-dimensional quantity when it is in reality at least a two-dimensional quantity, i.e., there are multiple different heights for the door and the door frame, varying across its breadth and width. Hence, the experienced carpenter would use the carpenter's square to ensure that the edges of the door are perpendicular to the length, to match the door frame, and reduce this uncertainty.

### c. Random error in the independent variable does not average away

The essence of regression analysis is to find the average of the measurement $Y^*$ when the measurement $X^*$ is a specific value '$x$'. This is equivalent to the conditional expectation of $Y^*$ given $X^* = x$, mathematically represented as $\langle Y^* | X^* = x \rangle$. Here, $Y^*$ is the dependent variable and $X^*$ is the independent or conditional variable. The averaging of $Y^*$ indeed removes the effect of random error in $Y^*$, consistent with the widely held belief. But, crucially, the conditional or independent variable $X^*$ cannot be averaged while simultaneously averaging the dependent variable $Y^*$ for a given value of $X^*$. In other words, the random error in the conditional variable remains unaddressed and is not averaged away[17]. And, surprisingly, this unbiased random error in the conditional variable ($X^*$) manifests as a bias in the relation inferred between the variables $Y^*$ and $X^*$. This *regression bias* is such that the average value of $Y^*$ for a given $X^*$ will be closer to the mean of $Y$, and if the regression bias is non-linear, it can create an appearance of saturation in $Y^*$ for increasing $X^*$. This statistical phenomenon is a result of a *regression to the mean*[6,18].

### d. Regression to the mean from different vantage points

There are five vantage points from which we can explain the regression to the mean effect.

**Vantage point 1: Regression bias.** The average value of a dependent variable given the independent variable will be biased closer to the mean value when there is some measurement uncertainty in the independent variable. Explained in Supplementary Discussion 1.c.

**Vantage point 2: Repeat measurements.** Extreme measurements are more likely followed by measurements closer to the mean. This is the most common understanding of the regression to the mean effect.

**Vantage point 3: Likelihood of Stochastic Processes.** Any physical process has an underlying probability distribution; therefore, some values of the physical process are more likely than others. For a stochastic process with a Gaussian distribution, the mean value coincides with the most likely or probable value. As a result, any uncertain measurement of the stochastic process is more likely measuring the true value closer to the mean of the stochastic process than more extreme values, by the simple fact that values closer to the mean are more common and hence more likely than the extreme values. In other words, the true value being measured is more likely to be closer to the mean of the stochastic process than the measurement.

**Vantage point 4: Probability Theory.** Extended Data Fig. 1a geometrically shows that for a Gaussian process, the true value is likely closer to the mean than the measurement. The black line is the probability distribution of a Gaussian process, X. When this process is measured with some uncertainty, the measurement can take the values in the likelihood quantified by the dashed-blue curve when the true value is X=2 and the dashed-red curve when it is X=4. Since the true value is generally inaccessible, the goal of a measurement is to estimate the most probable true value corresponding to the measurement. Hence, when the measurement X*=3 (solid black vertical line), the probability that the corresponding true value X is also equal to 3 (black dot) is remarkably *less likely* than it being equal to 2 (blue dot), simply because X=3 is less likely to occur than X=2 for a Gaussian process with zero mean.

**Vantage point 5: Relation between truth and measurement.** From vantage point 4, and consistent with other three vantage points, we can infer that regression to the mean is a property of the relation between truth and measurement, the effects of which are observable in statistical methods, and can be examined through probability theory. This new vantage point is the general result presented in this manuscript, further explanation is provided in Supplementary Discussion 1.e. The fact that regression to the mean is not only a statistical effect, but also a property of the relation between truth and measurement, imply that the effect is present even where a single measurement is made, and hence is likely ubiquitous.

### e. Regression to the mean as a fundamental property of the relation between truth and measurement

By fundamental property we mean a property that exists in reality, independent of the methodology we use to study it, as opposed to an effect that arises from the methodology. For example, an undersampling bias is an effect that arises purely from the number of samples we use in our analysis. All methods that use sufficient samples will never know of the undersampling bias. We argue that the regression to the mean effect, on the contrary, exists in reality as a property of the relation between truth and measurement independent of the methodology we use.

The first evidence for this comes from the fact that the phenomenon is observable in multiple statistical methods that use populations with inherent variability (e.g., different types of regression analysis, repeat measurements, machine learning, estimating conditional probability distributions), and also observable through probability theory when inferring the truth corresponding to the measurement. Supplementary Discussion 1.d explains regression to the mean effect from different vantage points. The second evidence is the mathematical result that can be derived using probability theory, graphically illustrated using Extended Data Fig. 1a, which shows that the effect only depends on the nature of the truth and measurement i.e., their likelihood of occurrence and their uncertainty. This implies that even when one makes a single measurement, the regression to the mean effect leads to a bias in the true value corresponding to the measurement according to the stochastic properties of the phenomenon being measured and the uncertainty in the measurement.

Here, an example of a familiar but different fundamental property of relations between things, e.g., the electric field, might be instructive and help us see the features of a fundamental property. An electric field is a fundamental property of the relationship between two or more charges. This relation, like all relations, has a qualitive aspect and a quantitative aspect. The quality of electric field is that it displaces the two charges relative to each other. The quantitative aspect is that the displacement depends on the nature of both the charges, its sign, magnitude, and distance between them. It was demonstrated to exist through experiments conducted by Faraday, but that does not mean it is only an experimental effect. It was proven to be mathematically consistent by Maxwell, but that does not mean it is only a mathematical effect or “field-theoretic” effect. This relationship between charges exists in reality, irrespective of the tools we use to observe it.

Similarly, when we say regression to the mean or the more-probable, is a fundamental property of the relationship between truth and measurements, it also implies, this relation, like all relations, has a qualitative and quantitative aspect. The quality of this property is that the truth is separated from the measurement by some measure. The distance of separation depends on the nature of the truth and the measurement, i.e., the stochastic properties of the truth, the degree of uncertainty of the measurement, and how it changes with time, space and magnitude of the truth. It can be demonstrated to exist through experiments with a large number of measurements (i.e., statistical methods), but that does not mean it is only a statistical effect. It can be proven to be mathematically consistent using probability theory (Extended Data Fig. 1a), but that does not mean it is only a probabilistic effect. This relationship between truth and measurement exists in reality, irrespective of the tools we use to observe this relationship.

The existence of this relation between truth and measurement arises only because they are different from each other. Here we provide a philosophical insight into why that is the case. In our universe, truth is generally only accessible through measurements - this is because the universe is a complex system. Complex systems have many processes, that the most prominent process will have a systematic effect that we can observe, which we may call a “law”, or we infer as the physical process of our interest. However, on the flip side, it will have other processes that appear as random perturbations, which may not be of our interest, and hence appear as randomness. This forces a separation between truth and measurements, leading them to be both different things. Truth is not measurement, and measurement is not the truth. However, there is an inseparable connection between them. Truth cannot be accessed without measurements, and measurements have no independent existence without the truth. This firmly establishes the existence of at least one, if not many, relationships between truth and measurement. One such relation is the regression to the mean of the truth, where the randomness manifests as uncertainty in measurements, and stochastic variations in the truth, leading to a separation of the truth from measurement in quantifiable ways.

### f. Regression bias due to magnitude and time uncertainty

As shown by Sivadas & Sibeck, 2022[18] using a simple Monte-Carlo error model, if $X$ has a Gaussian distribution, and the error $\epsilon$ is also Gaussian and independent of $X$ like most instrumental errors (i.e., homoskedastic error[20]), then the likely true value for a given measurement (i.e., the average of $X$ given $X^*$ or $\langle X|X^*\rangle$) will have a linear bias. The purple curve in Extended Data Fig. 1b shows this linear bias as a lower slope than the dashed-black line of equality representing the curve if there were no uncertainty (i.e., the true unbiased relationship). However, suppose the error $\epsilon$ depends on $X$, and varies with the magnitude of $X$ (i.e., heteroskedastic error[20]) like some uncertainties related to the ‘problem of definition’, and $X$ has a log-normal distribution like many solar wind parameters[21]. In that case, the key point is that the likely true value for a given measurement will have a larger and non-linear bias that mimics a *saturation* of $X$ with increasing $X^*$. Extended Data Fig. 1c demonstrates this in the purple curve whose slope has a non-linear bias with respect to the black-dashed line of equality representing the true unbiased relationship. Supposing that $X$ is the true solar wind driver near the Earth, and $X^*$ is the uncertain measurement far from Earth, the response $Y$ of the Earth will be driven by the true driver $X$ and, crucially, not by the measurement $X^*$. If the response $Y$ is linearly related to $X$ (i.e., $Y \propto X$), then naturally, the response $Y$ will also appear to saturate with increasing value of the measurement $X^*$.

Furthermore, if we keep aside the magnitude uncertainty (i.e., set $\epsilon = 0$), and examine the effect of uncertainty in time $dt$, we observe that it will lead to a non-linear bias in the average of $X$ given $X^*$ as well, with increasing deviation from linearity with growing uncertainty in time defined by the ratio of the temporal uncertainty ($\delta$) and the autocorrelation time constant ($k$) of the stochastic process. Extended Data Fig. 4c shows quantitatively how the average of the true value $X$ given the measurement $X^*$ has a non-linear bias in its slope with respect to the dashed-black line of equality representing the true unbiased relationship. This non-linear bias increases with increasing ratio of $\delta/k$ , a dimensionless measure of the uncertainty in time. In the Supplementary Methods 2.c, we derive analytically from first principles this non-linear bias and its dependence on temporal uncertainty, which will be relevant for any study with measurement of systems that have a delay in their drivers or response, due to propagation of information, or dynamics of the system e.g., seismic waves from earthquakes[73], or response of medical patients to a specific chronic pain treatment[24].

### g. Interpretation of the Error Model output

Once the calculated uncertainty values are worked into the error model, and $X$ is assumed to have a log-normal distribution with an autocorrelation coefficient of ~100 minutes observed in solar wind measurements, the Monte-Carlo error model solves for $X^*$ and predicts an appearance of saturation of $X$ given $X^*$ (See Figure 2b). This saturated curve, and the conditional probability distribution of $X$ given $X^*$, matches surprisingly well with the saturation that appears in the data for the polar cap index PCI given the merging electric field estimated from solar wind measurements $E_m^*$ (See Figure 2a). The solution of the error model is consistent with data as viewed through both the conditional expectation $\langle PCI|E_m^* \rangle$ and the conditional probability distribution of $pdf(PCI|E_m^*)$ (See Figure 2a-b). Further validation of the error model with second order statistics of data is presented in the Methods section[19] and Supplementary Methods 2.b. Here $PCI$ is analogous to $Y$ in the error model, $E_m^*$ is analogous to $X^*$, and $E_m^{sh}$ analogous to $X$. The similarity in the curves $\langle PCI|E_m^* \rangle$ and $\langle X|X^* \rangle$, indicate that the geomagnetic response is proportional to the true shocked solar wind driver i.e., $PCI \propto E_m^{sh}$ and equivalently $Y \propto X$. The preconditions of this non-linear saturation relation between $PCI$ and $E_m^*$ are: 1) the heteroskedastic nature of the error in $E_m^*$, 2) the propagation time uncertainty, 3) the fact that the solar wind driver is a log-normal process with a particular autocorrelation time constant.

### h. A different measure of geomagnetic response is also linear

Our finding that uncertainties in the solar wind driver results in the saturation of the polar cap index, raises the possibility that other measures of the geomagnetic response previously reported to saturate (or not)[74] based on uncertain solar wind driver estimates may be incorrect. The SML index, which is the westward auroral electrojet strength, measured by completely different magnetometers located at high latitudes ($\sim 65^\circ N$ ) around the planet, also appears to saturate with erroneous solar wind driving estimates. However, the corrected solar wind driver values, calibrated exactly the same way as above, show a linear relationship with the westward auroral electrojet strength as well—a further confirmation that the high-latitude geomagnetic response does not saturate with solar wind driving, once we remove the effect of uncertainties. Figure 3b shows the linear relation of the SML index with corrected solar wind driver in the purple curve, once the regression calibration is applied to the erroneous data in the saturating green curve.

### i. Historical development of the polar cap potential saturation problem

The regression to the mean of extreme solar wind driver values can sometimes be confused with under-sampling bias. Under-sampling is a statistical effect entirely distinct from the regression to the mean effect. The under-sampling bias reduces with increasing samples, however, the regression to the mean remains the same even with infinite samples. It is easy to see this from the analytical derivation in the methods section and Supplementary Methods 2.c (See Equation 1.21), which is immune to any sampling bias. However, under-sampling has had an important impact on the development of the polar cap saturation problem[22].

When examining space science literature on this problem, we observe from the 1980s to the 2000s (See Extended Data Fig. 6a-b) that as more and more solar wind data accumulates, the polar cap potential's saturation limit trends upwards from ~0.5 mV/m to ~10 mV/m. As extreme values of solar wind drivers are rare, their samples in data analysis can be poor. The number of samples increases with time, and hence the bias of under-sampling also diminishes with time. As shown by Extended Data Fig. 6b, due to this bias, observations suggest different relations between solar wind driving and polar cap potential, with different saturation limits and even no saturation (i.e., Boyle et al., 1997[25]). Since inferences from data do not agree with each other, the developers of a theory of saturation are also confounded by the question of which observational study they ought to explain. Such a diversity of inferences from data itself confuses the development of theoretical explanations. Hence, it is not surprising that multiple theories have emerged over the years (See Extended Data Table 1) without a unifying theory explaining the saturation effect quantitatively, as there are multiple different quantitative relations between solar wind driving and polar cap potential depending on the study.

Our work, that carefully addresses uncertainties in the driver with a sufficiently large database of 25 years of measurements, brings a resolution to the contradiction between the different inferences and shows that the evidence overwhelmingly concludes that the relation between the driver and the response is linear. It is this linear relationship between the driver and response that a new (or old) physical theory or model needs to explain and validate their predictions against.

### j. Evidence in Literature

Some previous data-based studies like Boyle et al. (1997)[25] have found that solar wind driving varies linearly with cross-polar cap potential using spacecraft measurements after a meticulous filtering of data. Pulkkinen et al. 2016[75] found that on average, the auroral electrojet index varies linearly with the shocked solar wind electric field in the magnetosheath tangential to the magnetopause (a proxy for the local reconnection electric field). Additionally, they found that this magnetosheath electric field varies non-linearly (or saturates) with the solar wind electric field. Their findings are entirely consistent with our explanation that the variance of solar wind parameters relative to the local magnetosheath values is the source of the saturation effect and not some magnetosheath phenomenon. The increased variance observable within their own data suggests that the variance in the driver leads to the saturation, and not any physical effect that causes a consistent reduction in the local magnetosheath electric field as they suggest. Borovsky 2021[74] found that the ionospheric response of PCI and SML indices varies linearly with solar wind driving as quantified by the $S_{(1)(9b)}$ index. The $S_{(1)(9b)}$ index is developed using the method of canonical correlation analysis applied to a set of solar wind parameters and another set of geomagnetic responses. From many parameters in both sets, the method extracts a small number of canonical variate pairs that contain the majority of shared information, relegating the uncertainty (and variance) that causes a lack of correlation to an orthogonal variate pair. As we have shown, since the uncertainty leads to the saturation effect, by reducing the uncertainty in the new driver $S_{(1)(9b)}$, Borovsky 2021 implicitly removed the non-linear regression bias in the relation between $S_{(1)(9b)}$ and geomagnetic response parameters like PCI and SML index. Hence, they have a linear relationship through the entire range of the solar wind driver $S_{(1)(9b)}$, consistent with our result.

# 2 Supplementary Methods

### a. Kan-Lee electric field is an approximation of the true solar wind driver

The Kan-Lee electric field is an approximation of the component of the reconnection electric field ($E_R = V_{sh}B_{sh}\sin(\theta'/2)$) perpendicular to the geomagnetic field that creates the cross-polar cap electric field (which is equal to $V_{sh}B_{sh}\sin^2(\theta'/2)$ due to a geometrical argument presented first by Kan and Lee)[48]. The approximation assumes that the shear angle between the magnetosheath and magnetospheric magnetic field $\theta'$ is equal to $\theta_{sw}$. Furthermore, the approximation assumes $V_{sh}B_{sh} \cong V_{sw}B_{T,sw}$, which is true most of the time 1) as the solar wind velocity component tangential to the bowshock is negligible compared to the normal component, and 2) as the tangential component of the electric field across a normal MHD shock is conserved. This allows the Kan-Lee electric field ($E_m^*$) to estimate the cross polar electric field $E_{PC}$ with just solar wind parameters. Our examination of data shows that for electric field values <3mV/m, which is 95% of all available data points in 25 years, the Kan-Lee electric field ($E_m^*$) propagated (or time-shifted) to the polar cap ionosphere from the L1 point is linearly proportional to the cross polar electric field $E_{PC}$. This is well-documented in literature and is also observed for other forms of the solar wind coupling functions and polar cap index [4,50,76].

### b. Validation of the Error Model

As time and magnitude uncertainty increase, the regression bias also increases. A lower uncertainty than estimated by us will result in a decrease in regression bias. Therefore, a conservative estimate of both time and magnitude uncertainty is made by us to present a defensible case for our result.

For magnitude uncertainty, when $E_m^{sh}$ > 12 mV/m, we used uncertainties from simultaneous L1 measurements and solar wind values upstream of the bow shock, adopting a conservative estimate of the uncertainty (~30% relative error). For simultaneous magnetosheath measurements, the sample size is small, with fewer than 10 data points per bin, suggesting no extreme geomagnetic storms fall within the constraints of simultaneous L1 and sheath measurements near the subsolar point.

Similarly, for the time uncertainty $dt_1$, we use a scale parameter of ~8 minutes and a Student's t-distribution with shape factor 1.3 to simulate the narrow but heavy-tail distribution seen in Case & Wild's results[59]. Case & Wild estimate propagation delays directly by cross-correlating solar wind measurements from ACE and Cluster satellites, and compare these calculated delays with propagation delay estimates from the OMNI database we use. The regression bias caused by our estimated Student's t-distribution is slightly lower than the equivalent normal distribution with ~8 minutes $1\sigma$ standard deviation (i.e., it is a conservative estimate). Recently, Tasnim et al.[60] estimated the propagation delay using a combination of cross-correlation and validation methods, reporting a ~18-minute $2\sigma$ standard deviation in propagation delay uncertainty, consistent with our estimate.

For the time uncertainty $dt_2$, we use a standard deviation of ~25 minutes, consistent with the results of Milan et al.[16] and Stauning[51]. Milan et al.[16] derive the propagation delay to the ionosphere using cross-correlation of field-aligned currents and solar wind $B_z$. They find a time-lag distribution peaking at ~17 minutes with a broad peak ranging from 10-30 minutes. Stauning's[51] cross-correlation of PCN index with the solar wind merging electric field just upstream of the bow shock shows propagation delays from -5 to 50 minutes with an average of 20 minutes. We estimate the probability distribution of $dt_2$ as a Weibull distribution (shape factor 1.3) to capture the broad spread and ensure a positive total delay. The total regression bias caused by this distribution is again slightly lower than that of the equivalent normal distribution with a ~25-minute standard deviation, making it a conservative estimate. In other words, using a normal distribution for propagation time uncertainties would yield a similar regression bias as seen in our results, implying the results are insensitive to the heaviness of the distribution's tail.

Using the Monte Carlo simulation, we estimate that a 30% change in our estimated magnitude uncertainty will lead to only a ~12.18% change in the regression bias, with all other parameters held constant and dt=0. Using the analytical equation [1.21] in Supplementary Methods 2.c, we estimate that a 30% change in our estimated time uncertainty will lead to only a ~9.65% change in the regression bias, with all other parameters held constant and $\epsilon = 0$. Similarly, we estimate that a 30% change in our estimated standard deviation of X will lead to only a ~5.9% change in the regression bias given dt=33 minutes and $\epsilon = 0$. These numbers show that the regression bias predicted by the error model is insensitive to a 30% error in the model inputs, especially since the sensitivity is less than the minimum relative magnitude error of ~30% (See Figure 2c). Therefore, for a change in the input, we see a much smaller change in the error model's output. In other words, the error model is robust to change in the estimated input parameters. Note all these values are calculated for regression bias observed in X for a given X*=25mV/m, as it is the most extreme and rare measurement and has the maximum sensitivity of all the measurements.

To further validate our error model with the inputs we have estimated, we compare the statistics of the error model's outputs with those of the data.

- The model predicts the pdf of the erroneous estimate of the shocked solar wind driver $E_m^*$. As shown in Extended Data Fig. 3a, the pdf of $X^*$closely fits the pdf of $E_m^*$.

- The model predicts the standard deviation of the normalized error. A good way to visualize how the uncertainty varies with the magnitude of $X^*$ is to calculate the standard deviation of the normalized error $\sum(X^* - \mathrm{X})/\mathrm{X}$. This quantity largely matches with its counterpart in data $\sum(E_m^* - E_{PC})/E_{PC}$ (See Extended Data Fig. 3b). This match implies the shocked solar wind driver $E_m^{sh} \propto E_{PC}$, and our assumption is consistent with even the second order statistics. This also aligns with the fact that $E_{PC}$ is constructed by maximizing the correlation with $E_m^*$[3], hence all the contribution to the variance comes from the difference between $E_m^{sh}$ and $E_m^*$, instead of uncertainties in the measurement of $E_{PC}$. If the values we assume for the uncertainties $dt_1$, $dt_2$, and $\epsilon$, are inadequate, the model will underestimate or overestimate the statistical variation in the difference between $E_m^*$ and $E_{PC}$ shown in Extended Data Fig. 3b.

- Furthermore, the model predicts the conditional normalized error distribution itself. A more detailed picture of the statistical properties of $E_m^*$ can be obtained by plotting the conditional probability density of the normalized error given X or $E_{PC}$. In Extended Data Fig. 3d-e, the left plot is the conditional pdf of normalized error calculated from the data, and the plot on the right is the same from the model. Both have a similar structure, with the spread in error increasing up to ~12 mV/m and then decreasing. This is consistent with our estimate of magnitude uncertainty in Figure 2c.

- Finally, the model shows that the estimates of random, unbiased uncertainties in the solar wind input ($X^*$) lead to a conditional bias in the estimate of the true driver $X$, with the bias varying with the magnitude of $X^*$. This bias exactly reproduces that seen in the data. In Extended Data Fig. 3c, the black line is plotted from data and shows the conditional bias in the erroneous estimate $E_m^*$ given its strength. It matches remarkably with that calculated by the model – shown in purple. This non-linearly increasing bias is ultimately a result of the regression to the mean effect[6,18,20].

Extended Data Fig. 5 is another way to explain this effect. The left panel shows the probability distribution of the true value $X$. Each bin associated with X is assigned a unique color. The panel on the right shows the pdf of the error-prone estimate of X, i.e., $X^*$. However, the colors still preserve their correspondence to the true value. The share of color in a particular $X^*$ bin describes the proportion of X values misidentified as $X^*$. In each $X^*$ bin, a larger proportion of lower X values are misidentified as $X^*$. In other words, for large $X^*$, X is less than expected. Converting this into the terminology of observations, we get back to the statement of the problem introduced in the main paper: for large values of $E_m^*$, $E_m^{sh}$ is less than expected or appears to saturate.

### c. Analytical derivation of non-linear regression bias due to uncertainty in time

Here we analytically derive the result that temporal uncertainty in measurement can lead to a perception of non-linearity in a linear system's correlation response. For this, we first assume a linear system with an input random variable $X$ and output response variable $Y$, such that

$$Y = X \quad [1.3]$$

The input is measured or estimated with a temporal uncertainty represented by the random variable $\Delta$. The erroneous estimate is $W$ and is related to the true input $X$ as

$$W = X(t + \Delta) \quad [1.4]$$

where the statistics of $X$and $\Delta$ are known

$$X \sim f_X(x)$$
$$\Delta \sim f_\Delta(\delta)$$

Our goal here is to find the erroneous regression function $\langle Y|W\rangle = f_r^*(w)$ and see how different it is from the true regression function $\langle Y|X\rangle$. Based on equation [1.3], $\langle Y|X\rangle$ is simply $f_r(x) = x$. We also note that since the system has a simple linear response $\langle Y|W\rangle = \langle X|W\rangle = f_r^*(w)$ . For the rest of the section, our goal is to derive the function $\langle X|W\rangle$ or $f_r^*(w)$.

Firstly, we note that since $W(t)$ is a stochastic process, from equation [1.4], the conditional distribution

$$f_{W|\Delta}(w|\Delta = \delta) = f_{X(t+\delta)}(w)$$

But since $X$ is assumed to be a stationary process,

$$f_X(w) = f_{X(t+\delta)}(w)$$
$$\Rightarrow f_{W|\Delta}(w|\Delta = \delta) = f_X(w) \quad [1.5]$$

However,

$$f_W(w) = \int f_{W,\Delta}(w,\delta)d\delta$$
$$= \int f_{W|\Delta}(w|\Delta = \delta) f_\Delta(\delta)d\delta \quad [1.6]$$

And from equation [1.6],

$$f_W(w) = \int f_X(w) f_\Delta(\delta)d\delta$$

As we assume $X$ and $\Delta$ are independent, and $\int f_\Delta(\delta)d\delta = 1$, hence

$$f_W(w) = f_X(w) \quad [1.7]$$

That is, **the marginal distribution of $W$ and $X$ has the same functional form**.

From equations [1.5] and [1.7], it follows that $W$ **and $\Delta$ are also independent**.

$$\begin{aligned} f_{W,\Delta}(w,\delta) &= f_{W|\Delta}(w|\delta)f_\Delta(\delta) \\ &= f_X(w)f_\Delta(\delta) \\ &= f_W(w)f_\Delta(\delta) \end{aligned} \quad [1.8]$$

Now we estimate the joint probability distribution function of $W$, $X$, and $\Delta$.

$$f_{W,X,\Delta}(w,x,\delta) = f_{W|X,\Delta}(w|x,\delta)f_{X,\Delta}(x,\delta)$$

Since $X$and $\Delta$are independent,

$$f_{W,X,\Delta}(w,x,\delta) = f_{W|X,\Delta}(w|x,\delta)f_X(x)f_\Delta(\delta) \quad [1.9]$$

By integrating the above over the range of $\Delta$, we can calculate the marginal pdf of $W$ and $X$.

$$f_{W,X}(w,x) = \int f_{W|X,\Delta}(w|x,\delta)f_X(x)f_\Delta(\delta)d\delta \quad [1.10]$$

Finally, we can calculate the conditional pdf of $X$ given $W$.

$$f_{X|W}(x|w) = \frac{f_{W,X}(w,x)}{f_W(w)}$$

From equation [1.7], we see that

$$f_{X|W}(x|w) = \frac{f_{W,X}(w,x)}{f_X(w)}$$

Using this and equation [1.10], we get

$$f_{X|W}(x|w) = \int f_{W|X,\Delta}(w|x,\delta)\frac{f_X(x)}{f_X(w)}f_\Delta(\delta)d\delta \quad [1.11]$$

Hence from the definition of expectations we get,

$$\langle X|W\rangle = f_r^*(w) = \int x f_{X|W}(x|W=w)dx \quad [1.12]$$

To solve equation [1.12], we need to assume a functional form for the pdf of $X$ and $\Delta$. $f_X(x)$ is assumed to be lognormal distribution, and $f_\Delta(\delta)$ to be a normal distribution. A lognormal random variable $X$ has a corresponding normally-distributed random variable $Z$ such that

$$\begin{gathered} Z \sim \phi(z, \mu_Z, \sigma_Z) \\ Z = \log X \\ \mu_Z = \log\left(\frac{\mu_X^2}{\sqrt{\mu_X^2 + \sigma_X^2}}\right) \\ \sigma_Z = \sqrt{\log\left(1 + \frac{\sigma_X^2}{\mu_X^2}\right)} \end{gathered} \quad [1.13]$$

Here, the function $\phi$ is the general normal distribution.

$$\phi(x, \mu, \sigma) = \frac{1}{\sqrt{2\pi}\sigma} e^{-(x-\mu)^2/2\sigma^2} \quad [1.14]$$

Therefore,

$$f_X(x) = \frac{1}{x}\phi(\log x, \mu_Z, \sigma_Z) \quad [1.15]$$

Additionally, as $X(t)$ is a stochastic process, the random variable at each time instance correlates to some degree with adjacent time instances. We define this using an autocorrelation function of the following form

$$\rho_X(\delta, k) = e^{-\delta/k} \quad [1.16]$$

Here $\delta = |t_2 - t_1|$ is the absolute difference in time between two random variables, $X(t_2)$ and $X(t_1)$. For our case, this is also the error in measurement time corresponding to $\Delta \sim f_\Delta(\delta)$. ***k* is the autocorrelation time constant**.

The normally distributed uncertainty in time is defined as

$$f_\Delta(\delta) = \phi(\delta, 0, \sigma_\delta) \quad [1.17]$$

Here $\sigma_\delta$ **is a measure of the random uncertainty in the time of measurement** $\Delta$.

From equations [1.13] and [1.16], we can estimate the autocorrelation function corresponding to the normal random variable $Z$

$$\rho_Z(\delta, k) = \frac{1}{\sigma_Z^2}\log\left(1 + \rho_X(\delta, k)\left[e^{\sigma_Z^2} - 1\right]\right) \quad [1.18]$$

From the results of the bivariate lognormal distribution, we can derive the functional form of $f_{W|X,\Delta}(w|x,\delta)$ to be the following

$$f_{W|X,\Delta}(w|x,\delta) = \frac{1}{w}\phi\left(\log w, \mu_Z + \rho_Z(\delta,k)[\log x - \mu_Z], \sigma_Z\sqrt{1-\rho_Z(\delta,k)^2}\right) \quad [1.19]$$

From equations [1.11],[1.12],[1.15],[1.17], and [1.19], we can calculate the regression function$\langle X|W\rangle = f_r^*(w)$.

$$f_r^*(w) = \int_0^{\infty}\int_{-\infty}^{\infty} \frac{\phi\left(\log w, \mu_Z + \rho_Z(\delta,k)[\log x - \mu_Z], \sigma_Z\sqrt{1-\rho_Z(\delta,k)^2}\right)}{\frac{\phi(\log x, \mu_Z, \sigma_Z)}{\phi(\log w, \mu_Z, \sigma_Z)}\phi(\delta, 0, \sigma_\delta)} d\delta dx \quad [1.20]$$

By integrating the above integral within the ranges of $x$, and substituting $\frac{\delta}{k} \to u$, we can show that the above integral is a function of the ratio $\frac{\sigma_\delta}{k}$. This ratio as a measure of uncertainty in the time of measurement.

$$f_r^*\left(w,\frac{\sigma_\delta}{k}\right) = \int_{-\infty}^{\infty} w^{\rho_Z(u)} e^{\frac{1}{2}[1-\rho_Z(u)][2m_Z+(1+\rho_Z(u))\sigma_Z^2]} \frac{1}{\sqrt{2\pi}\,(\sigma_\delta/k)} e^{-\frac{u^2}{2(\sigma_\delta/k)^2}} du$$

$$where\ u = \delta/k \quad [1.21]$$

We numerically integrate the above integral within the ranges of $\Delta$ for specific values of $\sigma_\delta/k$ and generate Extended Data Fig. 4c. The pdf of $X$ is similar to that assumed in the previous section, with the mean $\mu_X = 1.12$ and standard deviation $\sigma_X = 1.15$ The figure shows that $f_r^*(w,\frac{\sigma_\delta}{k})$ varies non-linearly with $w$ when the temporal uncertainty ratio is larger than 0%, compared to $f_r(x) = x$ (which is linear) when $W = X$ or the temporal uncertainty ratio is 0%. Therefore, we have found that the regression function $\langle Y|W\rangle = f_r^*(w,\frac{\sigma_\delta}{k})$ is non-linear with respect to $w$ for values of $\frac{\sigma_\delta}{k} > 0$. For low values of the temporal uncertainty ratio $\frac{\sigma_\delta}{k} < \sim 0.1$, the regression bias is approximately linear. Finally, the numerical integration of the analytical solution described by [1.21] was compared with an independent Monte Carlo solution of the error model stated in equation [1.4], with the results being almost identical, confirming that our derivation is correct.

### d. Summary of data analysis procedure in the computer code

The primary computer code (Code 1) constructs a Monte-Carlo simulation to solve a non-linear error model to show that the polar cap index saturation observed in data is a result of the regression to mean effect stemming from random error in the shocked solar wind driver estimates.

**Step 1: Preparing and loading observations data.** First 1-minute resolution WIND spacecraft data already time-shifted to bow-shock nose from the OMNI database, Polar Cap Index (PCI) data, and the magnitude uncertainty (in the form of standard deviation) varying with shocked solar wind driver (derived in Code 3) is loaded. All missing values, and out-of-range values, are converted into a ‘nan’ value, PCI is calculated from PCN and PCS as described in the Methods section. The Kan-Lee electric field at L1 ($E_m^*$) is calculated from WIND spacecraft measurements of plasma and field parameters. Finally, all values of $E_m^*$ and PCI that are not simultaneous are removed from the analysis. The difference between $E_m^*$ and PCI from data is also calculated as a measure of the total error or difference between the driver and the response. Finally, the autocorrelation function of $E_m^*$ , PCI, and $E_m^* -$ PCI is calculated.

**Step 2: Constructing model input.** We compare the pdf of $E_m^*$ and PCI, create a log-normal pdf that best fits the PCI index values, which has mean 1.114 and standard deviation 1.14 and, in the log-space, has mean -0.2518 and standard deviation 0.85. We fit a spline-function on the autocorrelation values and compare the autocorrelation functions of $E_m^*$ , PCI, and $E_m^* -$ PCI. From these values derived from measurements, we create the first input to the model: a stochastic process $X$ which is the hypothetical true stochastic process (i.e., shocked solar wind driver value), with also an autocorrelation functions that is similar to that observed in data. For this we use an external function by Lienhard (2021) to generate multivariate lognormal random numbers with correlation. Then we construct random variables representing time uncertainties: dt1 and dt2 with distributions and parameters derived from previous literature and discussed in the Methods section. Next, we construct a stochastic process $\epsilon(t)$ representing the magnitude uncertainty, which has zero mean, and a variance varying with the magnitude of $X$ according to the uncertainty derived directly from simultaneous magnetosheath and solar wind measurements using Code 3. $\epsilon(t)$ also has an autocorrelation function with values similar to that derived from the autocorrelation function of $E_m^* - PCI$.

**Step 3: Solving the Monte-Carlo Error Model.** By plugging in the above inputs of $X, dt_1, dt_2$ and $\epsilon$, into the following error model $X^* = X\ (t + dt_1 + dt_2) + \epsilon$, we estimate the stochastic process $X^*$. These stochastic processes are represented by an ensemble of 16000 realizations, with each realization consisting of $2^{12}$ time samples (i.e., ~4096 minutes). This can be increased to improve the statistics at the very low probability tail region of the log-normal distribution (i.e., for solar wind driver values > $15\ mV/m^2$). The resulting value of $X^*$ is the solution of the error model. From this we calculate the conditional expectation of $\langle X|X^*\rangle$, and notice that there is an appearance of a saturation effect on the average shocked solar wind driver given solar wind driver values measured at L1. This is then compared with the conditional expectation of $\langle PCI|E_m^*\rangle$, and we notice that the curves have a good correspondence, suggesting the regression to the mean effect is the cause for the saturation of the polar cap index with solar wind driver value measured at L1.

**Step 3: Validating the Monte-Carlo Error Model.** We verify the reliability of the error model by comparing many second-order statistics, as well as full probability distribution and conditional probability distributions with that of their counter parts in measurements. We list here the different statistics from the model that have been compared to show good correspondence with their counterparts in data: probability density function of input - $pdf(X)$, probability density function of output - $pdf(X^*)$, the autocorrelation function of the input - $\langle X(t)X(t+\tau)\rangle$, autocorrelation function of the output - $\langle X^*(t)X^*(t+\tau)\rangle$, normalized marginal error distribution - $pdf\left(\frac{X^*-X}{X}\right)$, the normalized marginal error distribution conditioned on $X$ - $pdf\left(\frac{X^{*}-X}{X}|X\right)$, the normalized marginal error distribution conditioned on $X^{*}$ - $pdf\left(\frac{X^{*}-X}{X}|X^{*}\right)$, standard deviation of normalized error - $\sigma(X^{*}-X)/X$, mean of error - $\langle X^{*}-X\rangle$ and finally the conditional $pdf(X|X^{*})$.

**Step 4: Saturation effect in another geomagnetic response parameter (SML).** We load the westward auroral electrojet index (SML) from SuperMag database, remove all the data gaps, and select only the data points that are simultaneous with the time-shifted Kan-lee electric field measured from L1. We then assume a counter part for SML in the model, $Y_{SML} = -120X + \epsilon_{Y}$, where $\epsilon_{Y}$ is a Gaussian-distributed random error in the estimate of the SML value with zero mean and $\sigma_{\epsilon_{Y}} = 5$ units. When the conditional expectation of $Y_{SML}$ given $X^{*}$, $\langle Y_{SML}|X^{*}\rangle$, is calculated we observe that it also saturates very similar to the saturation of SML index with solar wind driver values.

**Step 5: Regression calibration: Correcting for the regression to the mean effect.** The bias caused by the regression to the mean effect is quantified by $b = X^{*} - \langle X|X^{*}\rangle$, and this can be removed from $X^{*}$ to estimate the corrected solar wind driver function $X^{c} = X^{*} - b = \langle X|X^{*}\rangle$. This is carried out in the code by a one-dimensional linear interpolation of the solar wind driver value measured at L1 and time-shifted to the bow-shock ($E_{m}^{*}$) with a look up table consisting of sample points $x_{i} \in [0,25]$ and its corresponding values $f(x_{i}) = \langle X|X^{*} = x_{i}\rangle$ for each query point $E_{m}^{*}$. Therefore, the result is $E_{m}^{c} = f_{r}(E_{m}^{*})$. We now plot the conditional expectation $\langle PCI|E_{m}^{c}\rangle$ and $\langle SML|E_{m}^{c}\rangle$ all derived directly from data. They reveal a linear relation of $PCI$ and $SML$ with the corrected solar wind driver function (or the true shocked solar wind driver value). We refer interested readers to the following literature for further discussion on the distributions of solar wind parameters, and geomagnetic response measures[17,18,74,77,78].

Code 2 integrates the analytical formula in Eq [1.21] to solve for the regression bias stemming from uncertainty in the time of the measurement of a stochastic process. The code outputs a sensitivity analysis of the regression bias for different values of the ratio of time uncertainty to the autocorrelation time constant. Lastly, we validate the analytical solution independently; by developing an error model of only time uncertainty and solving it using the Monte Carlo method. The sensitivity analysis from the analytical result and the Monte Carlo solution match very well, confirming our analytical derivation is correct.

Code 3 uses simultaneous magnetosheath satellite measurements and measurements of spacecraft at L1 from the OMNI database, to calculate the variance of the L1 solar wind driver estimates for a given value of the shocked solar wind driver close to magnetopause subsolar point. The resulting variance is the magnitude uncertainty, and a spline fit of the variation of the magnitude uncertainty with increasing value of the shocked solar wind driver is used in Code 1 to reconstruct the magnitude uncertainty $\varepsilon$ and its heteroskedastic variation with the magnitude of $X$ in the error model in Code 3.

Code 4 uses probability theory to show that the true value corresponding to an uncertain measurement of a Gaussian random process is biased towards the mean. This is a demonstration of the regression to the mean effect through the lens of probability theory.

# Additional References

69. Cowley, S. ~W. ~H. & Lockwood, M. Excitation and decay of solar wind-driven flows in the magnetosphere-ionosphere system. *Ann. Geophys.* **10**, 103–115 (1992).

70. Lockwood, M. Solar Wind—Magnetosphere Coupling Functions: Pitfalls, Limitations, and Applications. *Space Weather* **20**, e2021SW002989 (2022).

71. Troshichev, O. A. & Andrezen, V. G. The relationship between interplanetary quantities and magnetic activity in the southern polar cap. *Planet. Space Sci.* **33**, (1985).

72. Dunham, D. ~W. ISEE-3, the first libration-point satellite. in *Bulletin of the American Astronomical Society* vol. 11 805–806 (1979).

73. Bonilla, M. G., Mark, R. K. & Lienkaemper, J. J. Statistical relations among earthquake magnitude, surface rupture length, and surface fault displacement. *Bulletin of the Seismological Society of America* **74**, 2379–2411 (1984).

74. Borovsky, J. E. On the Saturation (or Not) of Geomagnetic Indices. *Frontiers in Astronomy and Space Sciences* **8**, 175 (2021).

75. Pulkkinen, T. I. *et al.* Magnetosheath control of solar wind-magnetosphere coupling efficiency. *J. Geophys. Res. Space Phys.* **121**, 8728–8739 (2016).

76. Doyle, M. A. & Burke, W. J. S3-2 measurements of the polar cap potential. *J. Geophys. Res. Space Phys.* **88**, 9125–9133 (1983).

77. Vörös, Z. *et al.* Probability density functions for the variable solar wind near the solar cycle minimum. *J. Geophys. Res. Space Phys.* **120**, 6152–6166 (2015).

78. Borovsky, J. E. Noise and Solar-Wind/Magnetosphere Coupling Studies: Data. *Frontiers in Astronomy and Space Sciences* **0**, (2022).